\documentclass[12pt]{article}
\pdfoutput=1 
\usepackage{amsmath,amssymb,amsfonts,amsthm}
\usepackage{graphicx}
\usepackage{xcolor, colortbl}
\usepackage{empheq}
\usepackage[pdftex, bookmarks=true,colorlinks,linkcolor=red,urlcolor=blue, citecolor=blue]{hyperref}
\usepackage{cite}
\usepackage{subcaption}
\usepackage{slashed}

\usepackage{tikz}
\usetikzlibrary{calc,arrows.meta,decorations.markings}
\usepackage{amsthm}
\usepackage{array}
\usepackage{soul}
\usepackage{float}
\usepackage{bm}

\makeatletter\@addtoreset{equation}{section}\makeatother

\newcommand{\preprint}[1]{\begin{table}[t]  
             \begin{flushright}               
             {#1}                             
             \end{flushright}                 
             \end{table}}                     
\renewcommand{\title}[1]{\vbox{\center\LARGE{#1}}\vspace{5mm}}
\renewcommand{\author}[1]{\vbox{\center#1}\vspace{5mm}}
\newcommand{\address}[1]{\vbox{\center\em#1}}

\newcommand{\be}{\begin{equation}}
\newcommand{\ee}{\end{equation}}
\newcommand{\bea}{\begin{eqnarray}}
\newcommand{\eea}{\end{eqnarray}}
\newcommand{\bse}{\begin{subequations}}
\newcommand{\ese}{\end{subequations}}
\newcommand{\beqa}{\begin{eqnarray}}
\newcommand{\eeqa}{\end{eqnarray}}
\newcommand{\beqar}{\begin{eqnarray*}}
\newcommand{\eeqar}{\end{eqnarray*}}
\newcommand{\bi}{\begin{itemize}}
\newcommand{\ei}{\end{itemize}}
\newcommand{\bn}{\begin{enumerate}}
\newcommand{\en}{\end{enumerate}}

\newcommand{\ba}{\begin{array}}
\newcommand{\ea}{\end{array}}
\newcommand{\bc}{\begin{center}}
\newcommand{\ec}{\end{center}}

\usepackage{amsmath,amssymb,amsfonts,bm}
\usepackage{mathtools}
\usepackage{physics}

\definecolor{darkgreen}{rgb}{0,0.3,0}
\definecolor{darkblue}{rgb}{0,0,0.3}
\definecolor{darkred}{rgb}{0.7,0,0}

\usepackage{tocloft}

\def\lae{\mathrel{\mathop{\smash{\lower .5 ex \hbox{$\stackrel<\sim$}}}}}
\def\lae{\mathrel{\mathop{\smash{\lower .5 ex \hbox{$\stackrel>\sim$}}}}}

\def\arXiv#1{\href{http://arxiv.org/abs/#1}{arXiv:#1}}
\def\arXiv#1#2{\href{http://arxiv.org/abs/#1}{arXiv:#1}}

\def\doi#1#2{\href{http://doi.org/#1}{#2}}

\begin{document}

\unitlength = .8mm

\hypersetup{pageanchor=false}
\begin{titlepage}
\vspace{.5cm}
\preprint{}

\date{\today}
\begin{center}
\hfill \\
\hfill \\
\vskip 1cm

\title{\boldmath Entanglement structure in   conformal 
Y-junctions 
}
\vskip 0.5cm
{Yan Liu$^{a, b}$}\footnote{Email: {\tt yanliu@buaa.edu.cn}} 
{and You-Jie Zeng$^a$}\footnote{Email: {\tt BY2330101@buaa.edu.cn}}

\address{$^{a}$Department of Space Science, Beihang University, Beijing 100191, China}

\address{$^{b}$Peng Huanwu Collaborative Center for Research and Education, \\ Beihang University, Beijing 100191, China}

\end{center}
\vskip 1.5cm


\begin{abstract}

We study how entanglement constrains energy transmission in conformal Y-junctions using holographic and free-field models. We define pairwise effective central charges from mutual information as a natural extension of the interface entanglement coefficient. For holographic junctions, we formulate a branched RT prescription and show that the resulting entropies satisfy monogamy of mutual information.  We find that the pairwise effective central charge does not always bound the corresponding energy transmission, but a correction involving finite tripartite information restores the bound
, with a sharp coefficient determined by limiting saturation. Free-boson and Dirac-fermion junctions also violate the pairwise bound, but the tripartite correction fails because the tripartite information is positive or vanishing, respectively. Analytic limits and numerical results instead support a candidate square-root correction based on L-entropy, which vanishes in the holographic complete-leg prescription considered here. We discuss the possibility of a common multipartite information bound in general conformal junctions. 
\end{abstract}

\end{titlepage}
\hypersetup{pageanchor=true}

\begingroup 
\hypersetup{linkcolor=black}
\setlength{\cftbeforesecskip}{9pt}
 \setlength{\cftbeforesubsecskip}{4pt}
\tableofcontents
\endgroup



\section{Introduction}

A conformal interface is a one-dimensional defect separating two conformal field theories (CFTs). In two spacetime dimensions, it can describe a scale-invariant impurity connecting two critical quantum wires. Energy transmission is encoded in the coefficient $c_{LR}$ of the cross-interface stress-tensor two-point function and can be probed by sending an energy pulse toward the interface~\cite{QuellaRunkelWatts2007,MeineriPenedonesRousset2020}. Entanglement across the interface is characterized by an effective central charge $c_{\mathrm{eff}}$, which determines the coefficient of the logarithmic term in the entanglement entropy of intervals adjacent to the interface~\cite{SakaiSatoh2008, Afxonidis:2025jph}. In particular, the bound $c_{\mathrm{eff}}\leq\min{c_L,c_R}$ was established in~\cite{KarchEtAlUniversality2023}. These results motivate the proposed relation between energy transmission and interface entanglement~\cite{Karch}:
\begin{equation}
0\leq c_{LR}\leq c_{\mathrm{eff}}\leq\min\{c_L,c_R\},
\label{eq:intro-interface-bound}
\end{equation}
where $c_L$ and $c_R$ are the central charges on the two sides.

A conformal junction generalizes this setup to several CFTs defined on half-lines that meet at a common endpoint. Under folding, the junction becomes a boundary condition for the product CFT, providing a framework for studying energy transport and entanglement~\cite{KimuraMurata2015,GM,CMC}. On the holographic side, conformal interfaces admit a thin-brane description in AdS$_3$~\cite{BachasEtAl2020}. Related developments have extended holographic interface constructions to smooth domain-wall geometries, including Janus interfaces, using thin-brane discretizations~\cite{BachasEtAlThick2023}. 
Here we focus on conformal Y-junctions at zero temperature. In our previous work, we constructed a holographic junction in which three AdS$_3$ regions meet along a common tensile brane and derived a ``one-versus-rest'' bound on the central-charge-weighted total transmission from one leg into the other two~\cite{LiuWangJunction2025}: $ c_i\sum_{j\ne i}\mathcal T_{ij}
\leq \min\{c_i,c_j+c_k\}
$~\cite{LiuWangJunction2025}.

For a pulse incident on leg $i$ of a Y-junction, energy can be reflected or transmitted into either of the other two legs. The one-versus-rest bound constrains the sum of the two transmitted channels, but does not relate either channel to the entanglement between the corresponding pair of legs. Since all three legs are coupled at the junction, the third leg can affect both the distribution of energy flux and the correlations between the two selected regions. This raises a natural question: can the entanglement between two selected legs bound the energy transmitted between them? To investigate this, we define a pairwise effective central charge $c_{ij}^{\mathrm{eff}}$ from the leading logarithm of mutual information between  disjoint regulated regions of a common macroscopic length adjacent to the junction.    Then we test whether the pairwise extension $c_i\mathcal T_{ij}\leq c_{ij}^{\mathrm{eff}}$ of the interface bound holds in a multi-party junction.

For the holographic junction, we formulate a branched RT prescription in which the homology regions on the three bulk branches have a common trace on the junction brane, building on previous developments 
\cite{RyuTakayanagi2006,LiuWangJunction2025, Guo:2025sbm, GuoMiaoNetwork2026}. We show that the resulting classical RT entropies satisfy subadditivity and monogamy of mutual information (MMI), implying nonpositive tripartite information $I_3$~\cite{HaydenHeadrickMaloney2013} for the disjoint regions considered here. We find violations of the pairwise bound in two families of holographic junctions and derive a corrected bound involving the UV-finite tripartite information:
$c_i\mathcal T_{ij}\leq c_{ij}^{\mathrm{eff}}+3(-I_3).$ 
MMI fixes the sign of $I_3$, while the coefficient $3$ is fixed by the asymptotic saturation of the corrected bound as $L_a\to\infty$ in the dominant-radius family. 

We then examine the pairwise bound in junctions of free real bosons and free Dirac fermions, described by a class of real symmetric orthogonal scattering matrices. Both theories provide counterexamples to the direct pairwise bound, but the holographic tripartite correction above does not provide a common remedy: in the bosonic realization studied here, the finite part of $I_3$ is positive, whereas it vanishes in the continuum Dirac example. 
This motivates considering a different measure of genuine tripartite correlations~\cite{BalasubramanianEtAl2024}. We use the L-entropy, constructed from reflected entropies for the partition of the pure ground state into the three complete legs~\cite{DF,LE}. Analytic limiting behavior and numerical calculations motivate a candidate square-root correction involving its logarithmic coefficient in the free-field models. By contrast, the L-entropy coefficient vanishes in the holographic complete-leg prescription considered here, so this correction cannot resolve the holographic pairwise-bound violations. Whether a common tripartite correction can apply to both settings remains an open problem. We comment on a possible resolution in the discussion section. 

The paper is organized as follows. In Sec.~\ref{sec:holographic-junction-review}, we review the holographic junction and its energy transmission coefficients. In Sec.~\ref{sec:entanglement-structure}, we develop the branched RT prescription, compute mutual and tripartite information, and establish the relevant entropy inequalities. In Sec.~\ref{sec:junction-transmission-bounds}, we examine the entropic constraints on energy transmission in holographic junctions. We further extend the analysis to free-field junctions in Sec.~\ref{sec:free-field-bounds}. We conclude in Sec.~\ref{sec:conclusion} with a summary of our findings and a discussion of possible multipartite corrections and open questions. In App.~\ref{app:junction-information}, we provide the detailed RT calculations of mutual and tripartite information.


\section{Review of holographic junctions}
\label{sec:holographic-junction-review}

We briefly review the holographic construction of conformal junctions and the corresponding energy transport coefficients, following~\cite{LiuWangJunction2025}. We first summarize the background geometry and its junction conditions, and then present the reflection and transmission coefficients obtained from linear perturbations. The construction applies to general $N$-leg junctions, although our main interest is the Y-junction.

\subsection{The junction and its gravitational description}
\label{subsec:hj-construction}

\begin{figure}[H]
\centering
\definecolor{cftnavy}{HTML}{244577}
\definecolor{junctionred}{HTML}{A82438}
\begin{tikzpicture}[
  font=\small, line cap=round,
  wire/.style={draw=cftnavy,line width=1.05pt},
  continuation/.style={wire,draw=cftnavy!55,
    dash pattern=on 1.5pt off 1.5pt},
  directed/.style={
    wire, decoration={markings,mark=at position .54 with
      {\arrow{Stealth[length=5.78pt,width=5.4pt]}}},
    postaction={decorate}
  },
  endpoint/.style={fill=junctionred},
]
  \coordinate (J) at (0,0);
  \foreach \i/\y in {1/1.42,2/.68,3/-.06,N/-1.42}{
    \coordinate (B\i) at (-2.45,\y);
    \draw[continuation] (-2.77,\y) -- (B\i);
    \draw[directed] (B\i) -- (J);
    \node[anchor=east,inner sep=0pt] at (-2.91,\y)
      {$\mathrm{CFT}_{\i}$};
  }
  \foreach \y in {-.44,-.65,-.86}
    \fill[cftnavy!65] (-2.45,\y) circle[radius=.85pt];
  \fill[endpoint] (J) circle[radius=2.65pt];
  \node[above=7pt] at (J) {$x_i=0$};

  \draw[-{Stealth[length=4.8pt,width=4pt]},line width=.7pt,blue!75]
    (.85,0) -- (2.38,0) node[midway,above=4pt] {\color{blue}folding};

  \coordinate (F) at (6.22,0);
  \draw[continuation] (2.94,0) -- (3.26,0);
  \draw[directed] (3.26,0) -- (F);
  \fill[endpoint] (F) circle[radius=2.65pt];
  \node[above=8pt] at (4.45,0)
    {$\displaystyle\bigotimes_{i=1}^{N}\mathrm{CFT}_i$};
  \node[below=4pt] at (4.45,0) {$x<0$};
  \node[above=7pt] at (F) {$x=0$};
  \node[below=4pt,align=center] at (F) {common\\boundary};

  \node at (-1.43,-1.94) {(a) Junction};
  \node at (4.58,-1.94) {(b) Folded description};
\end{tikzpicture}
\caption{\small A junction of $N$ CFTs (left) and its folded description (right). Folding places the product theory on a single half-line, with the junction encoded in a common boundary condition. Red points mark the junction. 
}
\label{fig:cft-junction}
\end{figure}
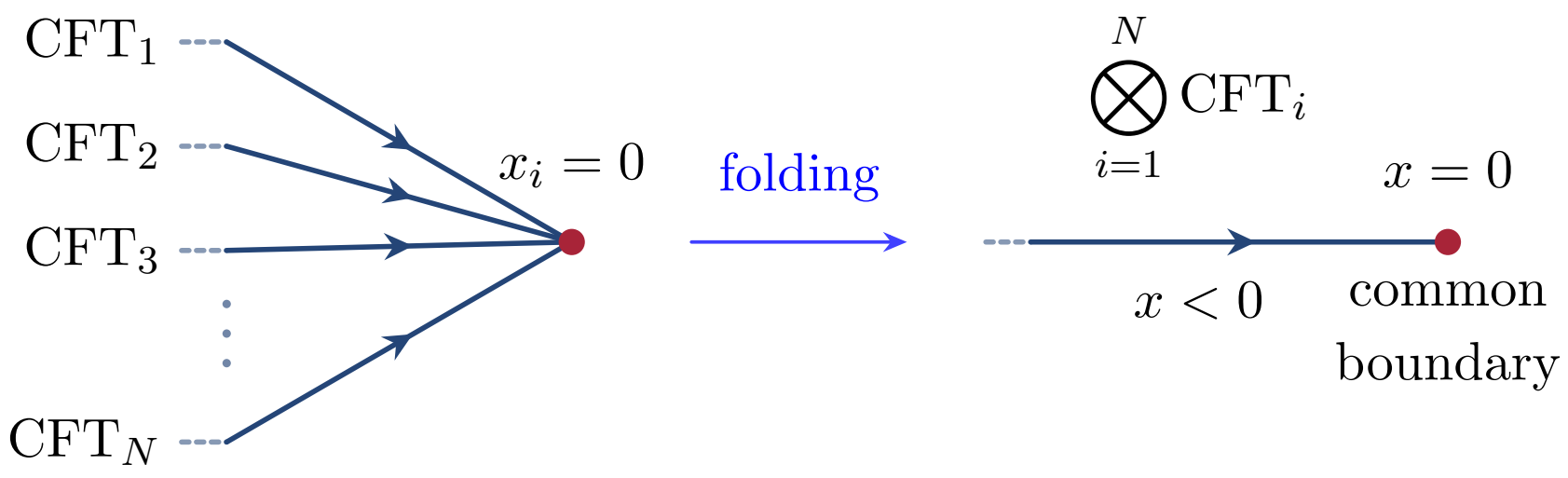

Consider $N$ two-dimensional CFTs with central charges $c_i$, each defined
on a half-line $x_i<0$. Their endpoints are coupled at $x_i=0$ to form a
junction, as shown in Fig.~\ref{fig:cft-junction}.
Folding maps the branches to a common half-line, with the junction becoming
a boundary of the product theory~\cite{KimuraMurata2015},
\begin{equation}
 \mathrm{CFT}_{\mathrm{fold}}=\bigotimes_{i=1}^{N}\mathrm{CFT}_i,
 \qquad c_{\mathrm{fold}}=\sum_{i=1}^{N}c_i.
 \label{eq:hj-folded-theory}
\end{equation}

The junction coupling is encoded in this boundary condition. Conformal
invariance requires 
\begin{equation}
 \left.\sum_{i=1}^{N}T_i\right|_{x=0}
 =\left.\sum_{i=1}^{N}\overline T_i\right|_{x=0},
 \label{eq:hj-conformal-boundary}
\end{equation}
where $T_i$ and $\overline T_i$ are the chiral stress tensors in the folded 
description. This condition conserves the total energy flux while allowing transmission between branches. For $N=2$, there is a single transmission channel, whereas for $N\geq3$, the transmitted energy can be distributed among multiple branches.

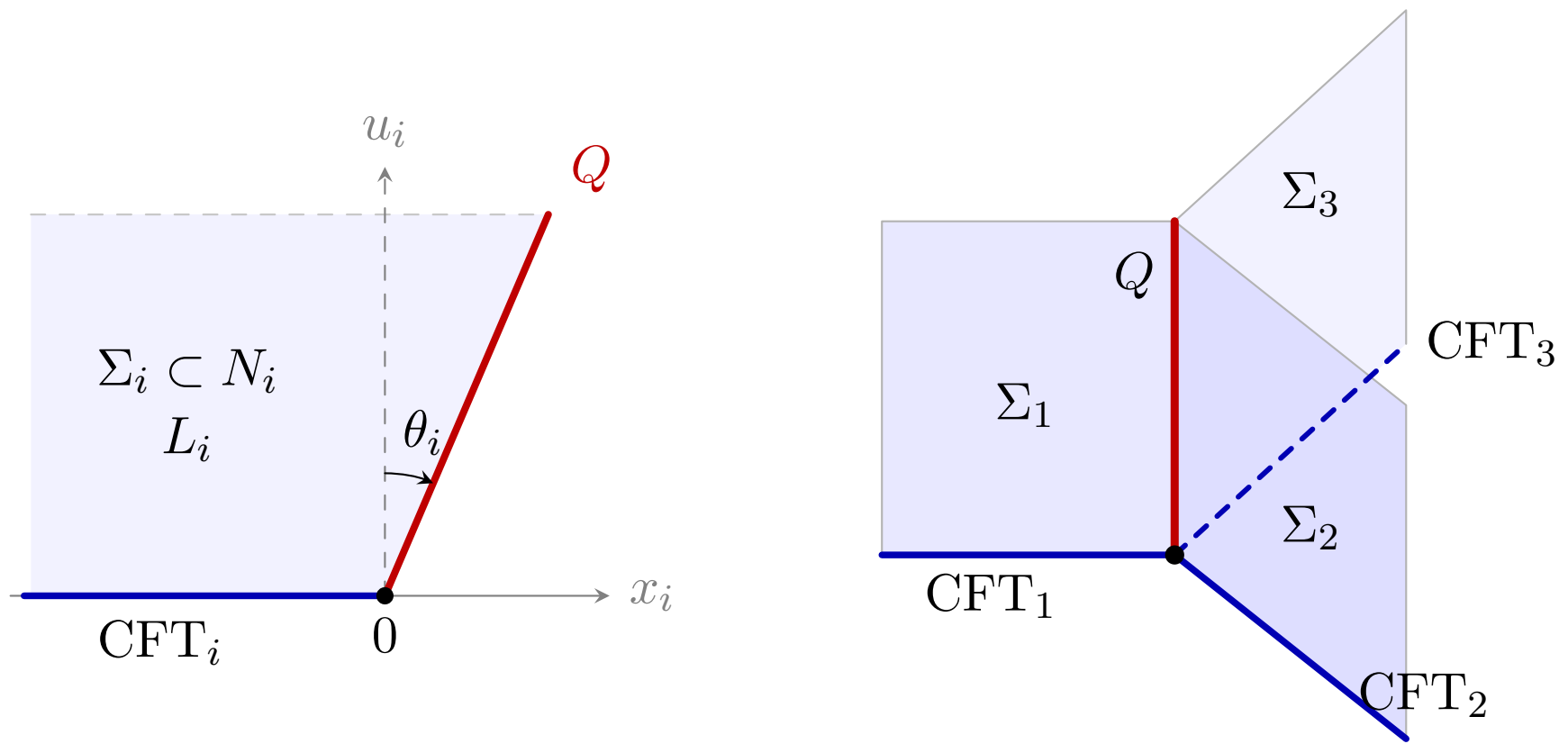
\begin{figure}[h!]
\centering
\begin{tikzpicture}[>=stealth,font=\small,line cap=round,line join=round]
\begin{scope}[shift={(2.8,0)}]
\fill[blue!5] (-2.6,0)--(0,0)--(1.2,2.8)--(-2.6,2.8)--cycle;
\draw[gray!45,dashed] (-2.6,2.8)--(1.2,2.8);
\draw[->,gray] (-2.75,0)--(1.65,0) node[right] {$x_i$};
\draw[->,gray,dashed] (0,0)--(0,3.15) node[above] {$u_i$};
\draw[blue!70!black,line width=1.4pt] (-2.65,0)--(0,0);
\draw[red!75!black,line width=1.5pt] (0,0)--(1.2,2.8) node[above right] {$Q$};
\draw[->] (0,.9) arc[start angle=90,end angle=66.8,radius=.9];
\node at (.28,1.2) {$\theta_i$};
\fill (0,0) circle (1.8pt);
\node[below] at (0,-.02) {$0$};
\node[below] at (-1.65,-.04) {$\mathrm{CFT}_i$};
\node at (-1.45,1.65) {$\Sigma_i\subset N_i$};
\node at (-1.45,1.15) {$L_i$};
\node at (-.65,-1.6) {(a) One bulk branch};
\end{scope}
\begin{scope}[shift={(8.6,.3)}]
\fill[blue!5] (0,0)--(1.7,1.55)--(1.7,4)--(0,2.45)--cycle;
\fill[blue!9] (-2.15,0)--(0,0)--(0,2.45)--(-2.15,2.45)--cycle;
\fill[blue!13] (0,0)--(1.7,-1.35)--(1.7,1.1)--(0,2.45)--cycle;
\draw[gray!60] (-2.15,0)--(-2.15,2.45)--(0,2.45)--(1.7,4)--(1.7,1.55);
\draw[gray!60] (0,2.45)--(1.7,1.1)--(1.7,-1.35);
\draw[blue!70!black,line width=1.4pt] (-2.15,0)--(0,0)--(1.7,-1.35);
\draw[blue!70!black,line width=1.1pt,dashed] (0,0)--(1.7,1.55);
\draw[red!75!black,line width=1.7pt] (0,0)--(0,2.45);
\fill (0,0) circle (2pt);
\node[left] at (0,2.05) {$Q$};
\node at (-1.1,1.1) {$\Sigma_1$};
\node at (1,.2) {$\Sigma_2$};
\node at (1,2.65) {$\Sigma_3$};
\node[below] at (-1.35,0) {$\mathrm{CFT}_1$};
\node[right] at (1.2,-1.03) {$\mathrm{CFT}_2$};
\node[right] at (1.7,1.55) {$\mathrm{CFT}_3$};
\node at (-.1,-1.9) {(b) A common junction};
\end{scope}
\end{tikzpicture}
\caption{\small Spatial geometry of the holographic junction. (a) A single AdS branch bounded by $Q$ (red) and the CFT half-line (blue). (b) Three bulk slices $\Sigma_i$ joined along the same $Q$. The opening angles in (b) are schematic.}
\label{fig:hj-geometry}
\end{figure}

In the holographic description, we associate each CFT$_i$ with an asymptotically AdS$_3$ region $N_i$ with radius $L_i$ and join these regions along a common timelike brane $Q$ anchored at the boundary junction~\cite{LiuWangJunction2025}. Fig.~\ref{fig:hj-geometry} illustrates the geometry on a constant-time slice.

We describe the junction by Einstein gravity in each bulk region, with a
common Newton constant $G_N$, coupled to a brane $Q$ of constant tension.
The corresponding action is
\begin{equation}
 I=\frac{1}{16\pi G_N}\sum_{i=1}^{N}
 \int_{N_i}d^3x\,\sqrt{-g_i}\left(R_i+\frac{2}{L_i^2}\right)
 +\frac{1}{8\pi G_N}\int_Qd^2y\,\sqrt{-h}
 \left(\sum_{i=1}^{N}K_i-\tau\right).
 \label{eq:hj-action}
\end{equation}
Here $h_{\mu\nu}$ is the induced metric on $Q$,  $K_i$ denotes the trace of the
extrinsic curvature computed with the normal pointing out of $N_i$, and $\tau$ is the brane tension. We leave asymptotic boundary terms and counterterms implicit. 


To glue the bulk regions, we require a common induced metric on $Q$,
\begin{equation}
 h_{\mu\nu}=g^{(i)}_{ab}\,
 \partial_\mu X_i^a\partial_\nu X_i^b\big|_Q,
 \qquad i=1,\ldots,N.
 \label{eq:hj-metric-matching}
\end{equation}
where $X_i^a(y)$ specifies the brane $Q$ (with coordinates $y^\mu$) embedding in $N_i$. 
Varying the action gives the bulk Einstein equations and the junction condition \cite{LiuWangJunction2025, Shen:2024itl, Shen:2024dun},
\begin{align}
 R^{(i)}_{ab}+\frac{2}{L_i^2}g^{(i)}_{ab}=0,
 ~~~~~
 \sum_i K^{(i)}_{\mu\nu}-\left(\sum_iK_i-\tau\right)h_{\mu\nu}&=0.
 \label{eq:hj-junction-condition}
\end{align}
Since $Q$ is two-dimensional, tracing this condition gives
$\sum_iK_i=2\tau$, reducing the junction condition to 
\begin{equation}
 \sum_i K^{(i)}_{\mu\nu}=\tau h_{\mu\nu}.
 \label{eq:hj-traced-condition}
\end{equation}

At zero temperature, we take each bulk region to be a portion of Poincare AdS$_3$,  
\begin{equation}
 ds_i^2=\frac{L_i^2}{u_i^2}
 \left(-dt_i^2+dx_i^2+du_i^2\right),
 \qquad x_i\leq u_i\tan\theta_i,
 \label{eq:hj-bulk-metric}
\end{equation}
where $-\pi/2<\theta_i<\pi/2$ and the equality specifies the brane location. We parametrize its embedding using common coordinates $(t,z)$ on $Q$,
\begin{equation}
 t_i=t,\qquad x_i=z\sin\theta_i,\qquad u_i=z\cos\theta_i.
 \label{eq:hj-embedding}
\end{equation}
The induced metric on $Q$ is AdS$_2$,
\begin{equation}
 ds_Q^2=\frac{L_a^2}{z^2}(-dt^2+dz^2),
 \qquad \frac{L_i}{\cos\theta_i}=L_a\quad\text{for every }i.
 \label{eq:hj-induced-radius}
\end{equation}
For this embedding, the extrinsic curvature computed with the
outward-pointing unit normal is
\begin{equation}
 K^{(i)}_{\mu\nu}=\frac{\sin\theta_i}{L_i}h_{\mu\nu}.
 \label{eq:hj-extrinsic-curvature}
\end{equation}
The junction condition then gives
\begin{equation}
 \tau=\sum_i\frac{\sin\theta_i}{L_i}.
 \label{eq:hj-tension}
\end{equation}
The central charge of each CFT is related to its bulk AdS radius by the
Brown--Henneaux formula~\cite{BrownHenneaux1986},
\begin{equation}
 c_i=\frac{3L_i}{2G_N}.
 \label{eq:hj-brown-henneaux}
\end{equation}

For $L_a\geq\max_iL_i$, we express the tension as
\begin{equation}
 \tau(L_a;\sigma_i)=\sum_i\frac{\sigma_i}{L_i}
 \sqrt{1-\frac{L_i^2}{L_a^2}},\qquad
 \sigma_i=\operatorname{sgn}(\sin\theta_i).
 \label{eq:hj-branches}
\end{equation}
Here $\sigma_i=\pm1$ denotes a brane tilted towards $\pm x_i$,
labeling different bulk configurations.
For the Y-junction, we focus on two regimes with a unique static bulk configuration at fixed radii and tension~\cite{LiuWangJunction2025}.
When $L_1\leq L_2<L_3$ with $L_1+L_2<L_3$, the allowed tension range is
\begin{equation}
 \frac{1}{L_1}+\frac{1}{L_2}-\frac{1}{L_3}
 <\tau<\frac{1}{L_1}+\frac{1}{L_2}+\frac{1}{L_3}.
 \label{eq:hj-unique-range}
\end{equation}
For $L_1\leq L_2=L_3$, we instead consider
\begin{equation}
 \frac{1}{L_1}\leq\tau<\frac{1}{L_1}+\frac{2}{L_2}.
 \label{eq:hj-equal-radius-range}
\end{equation}
In both cases, the upper endpoint is $\tau_{\max}=\sum_iL_i^{-1}$,
approached as $L_a\to\infty$ with all $\theta_i>0$.

\subsection{Reflection and transmission from linear perturbations}
\label{subsec:hj-transport}

We study energy transport by scattering a small perturbation incident
from CFT$_i$. The corresponding normalizable metric perturbations
take the form~
\cite{BachasEtAl2020,LiuWangJunction2025}
\begin{align}
 \delta ds_i^2={}&4G_NL_i\varepsilon
 \left[e^{i\omega(t_i-x_i)}(dt_i-dx_i)^2
 +\mathcal R_i e^{i\omega(t_i+x_i)}(dt_i+dx_i)^2\right]
 +\mathrm{c.c.},\label{eq:hj-incoming-wave}\\
 \delta ds_j^2={}&4G_NL_j\varepsilon\,
 \mathcal T_{ij}e^{i\omega(t_j+x_j)}(dt_j+dx_j)^2
 +\mathrm{c.c.},\qquad j\ne i.
 \label{eq:hj-transmitted-wave}
\end{align}
Since all $x_i$ increase towards the junction, the first term in eq.~\eqref{eq:hj-incoming-wave} describes the incoming wave, while $\mathcal R_i$ and
$\mathcal T_{ij}$ characterize the reflected and transmitted
waves, respectively.

To determine these coefficients, we include brane fluctuations
and solve the linearized metric-matching and junction conditions,
imposing infalling conditions at the Poincare horizons and
vanishing displacement sources at the AdS boundaries~\cite{LiuWangJunction2025}.
Energy conservation requires
\begin{equation}
 \mathcal R_i+\sum_{j\ne i}\mathcal T_{ij}=1,
 \label{eq:hj-energy-conservation}
\end{equation}
which equates the total outgoing flux to the incident flux.
Introducing
\begin{equation}
 \eta_i\equiv\frac{\cos\theta_i}{1+\sin\theta_i}
 =\sqrt{\frac{1-\sin\theta_i}{1+\sin\theta_i}},
 \qquad E\equiv\sum_{k=1}^{N}\eta_k,
 \label{eq:hj-weights}
\end{equation}
we obtain~\cite{LiuWangJunction2025}
\begin{align}
\begin{split}
 \mathcal T_{ij}&=\frac{2\eta_j}{(1+\sin\theta_i)E},\qquad j\ne i,\\
 \mathcal R_i&=1-\frac{2(E-\eta_i)}{(1+\sin\theta_i)E}
 =\frac{\eta_i}{E}\left[1-\eta_i(E-\eta_i)\right].
 \label{eq:hj-reflection}
 \end{split}
\end{align}
The coefficients are real and frequency-independent, as expected
for a conformal junction. For a fixed incident branch, the
transmitted fluxes are proportional to the weights $\eta_j$. 

To compare energy transmission with entanglement, we define
$c_a=3L_a/(2G_N)$. Using $c_i=c_a\cos\theta_i$, we obtain
the central-charge-weighted transmission coefficients
\begin{equation}
 c_{ij}\equiv c_i\mathcal T_{ij}
 =\frac{2c_a\eta_i\eta_j}{E},\qquad i\ne j.
 \label{eq:hj-weighted-transmission}
\end{equation}
This expression is symmetric under $i\leftrightarrow j$, yielding
\begin{equation}
 c_{ij}=c_{ji}.
 \label{eq:hj-reciprocity}
\end{equation}
Thus, although $\mathcal T_{ij}$ need not be symmetric,
the weighted transmission is reciprocal.

Since $\eta_i>0$, all transmission coefficients are positive. Equation~\eqref{eq:hj-reflection} further shows that
nonnegative reflection requires 
\begin{equation}
 \eta_i\sum_{j\ne i}\eta_j\leq1\qquad\text{for every }i.
 \label{eq:hj-positive-reflection}
\end{equation}
Together with energy conservation and reciprocity, this gives 
$0\leq\mathcal R_i,\mathcal T_{ij}\leq1$ and 
$0\leq c_{ij}\leq\min(c_i,c_j)$.

For a junction of $N$ identical CFTs, consider the symmetric configuration
with equal AdS radii and a common brane angle $\theta$ \cite{LiuWangJunction2025,  Chakraborty:2026wip}. The transport
coefficients reduce to
\begin{equation}
 \mathcal T_{ij}=\frac{2}{N(1+\sin\theta)},\qquad
 \mathcal R_i=1-\frac{2(N-1)}{N(1+\sin\theta)}.
 \label{eq:hj-symmetric}
\end{equation}
All outgoing branches receive equal transmitted flux. Nonnegative reflection
requires $\sin\theta\geq(N-2)/N$, which becomes $\sin\theta\geq1/3$
for a symmetric Y-junction.

Finally, we can recover the two-CFT interface by taking
$L_3\to0$ and $\theta_3\to\pi/2$ at fixed $L_a$.
In this limit, $c_3,\eta_3\to0$, and transmission
into the third branch vanishes. The remaining coefficient is
\begin{equation}
 \mathcal T_{12}=\frac{2\cos\theta_2}
 {\cos\theta_2(1+\sin\theta_1)+
  \cos\theta_1(1+\sin\theta_2)},
 \label{eq:hj-interface-limit}
\end{equation}
in agreement with the thin-brane interface result~\cite{BachasEtAl2020}.
The reverse transmission follows by exchanging $1$ and $2$, and
$\mathcal R_1=1-\mathcal T_{12}$, $\mathcal R_2=1-\mathcal T_{21}$.

\section{Entanglement structure in holographic junctions}
\label{sec:entanglement-structure}

Energy transmission describes the exchange of energy between the branches of a junction. To relate it to information quantities, we need an entropy prescription for general boundary subregions. For holographic junctions, we apply the standard RT prescription~\cite{RyuTakayanagi2006} to the junction geometry studied in~\cite{LiuWangJunction2025, GuoMiaoNetwork2026}. For multipartite regions, we formulate an explicit branched RT prescription by imposing a common-trace homology condition across the bulk branches. We use this prescription to compute mutual and tripartite information and establish the relevant entropy inequalities. Throughout this section, we work on a common static time slice and retain only the leading classical RT contribution.

\subsection{The branched RT prescription
}
\label{subsec:rt-networks}

All geometric constructions below are restricted to a constant-time slice. Throughout this discussion, we denote the three bulk branches and their common junction brane by $N_i$ and $Q$, respectively. 
For a boundary subregion $A$, we require an admissible RT candidate $\Gamma_A$ to bound, together with $A$, a bulk homology region $\gamma_A$. Its portions in different branches must join consistently across $Q$.



\begin{figure}[htbp]
\centering\includegraphics[width=\textwidth]{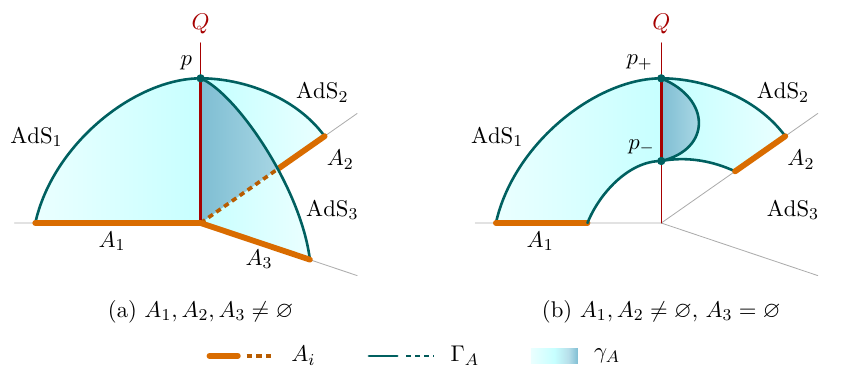}
\vspace{-0.8cm}
\caption{\small Possible branched RT configurations on a static slice, with $A_i=A\cap\mathrm{CFT}_i$. Orange intervals and teal RT curves bound the shaded spatial entanglement wedge after gluing the three bulk branches along $Q$. (a) All three boundaries contain subregions; the RT curves meet at $p$. (b) Only two boundaries contain subregions; a curve in the third bulk branch connects $p_-$ to $p_+$ and closes the wedge. All branches occupy the same segment of $Q$: from the boundary node to $p$ in (a), and $[p_-,p_+]$ in (b). These are illustrative candidates, not necessarily the minimizing configurations; curve shapes and vertex positions are schematic.}
\label{fig:ee-wedge}
\end{figure}

To glue the homology regions across the junction brane $Q$, the
portion in each bulk branch must occupy the same subset of $Q$.
Let $\gamma_{A,i}=\gamma_A\cap N_i$. Its characteristic function
$\chi_{A,i}$ equals one inside $\gamma_{A,i}$ and zero outside it.
Following the use of characteristic functions to represent bulk regions
in holographic entropy calculations\cite{HeadrickKasireddyRolph2026}, 
we impose {\em the common-trace 
condition}
\begin{equation}
 \left.\chi_{A,1}\right|_Q
 =\left.\chi_{A,2}\right|_Q
 =\left.\chi_{A,3}\right|_Q
 \equiv\chi_{A,Q}.
 \label{eq:ee-common-trace}
\end{equation}
At an interior endpoint $p$ of an occupied segment of $Q$, the
common trace jumps between one and zero. Since the jump occurs at
the same point in all three branches, the boundary of each
$\gamma_{A,i}$ reaches $Q$ at $p$.
At a regular vertex, the three boundary curves consequently form a Y-shaped RT surface. This realizes the branching of RT surfaces at the junction brane: the curves jointly bound a bulk region with the prescribed boundary subregion.  In particular, a curve cannot terminate freely on $Q$  without matching boundaries in the other branches.
Fig.~\ref{fig:ee-wedge} illustrates the possible occupied segment of the junction and the resulting branched RT surface.

The common-trace condition does not force every admissible RT configuration to intersect $Q$:  curves confined to a single bulk branch remain allowed when they satisfy homology. We also include general configurations containing disconnected components. For any $A$, we obtain the entropy by minimizing the total length over all candidate RT surfaces $\Gamma_A$ that, together with $A$, bound a bulk region $\gamma_A$ whose branch portions satisfy the gluing condition~\eqref{eq:ee-common-trace}:
\begin{equation}
 S(A)=\frac{1}{4G_N}\min_{\mathrm{admissible~} \Gamma_A}
 \mathcal L(\Gamma_A),
 \qquad\qquad
 \mathcal L(\Gamma_A)
 =\sum_{i=1}^{3}
 \operatorname{Length}\!\left(
 \partial \gamma_{A,i}\cap\operatorname{int}N_i\right).
 \label{eq:ee-prescription}
\end{equation}
Here only the bulk RT curves contribute to the length sum. Henceforth, $\Gamma_A$ and $\gamma_A$ refer to a minimizing configuration. 

For fixed junction vertices, the length minimization gives geodesic segments in each bulk branch. The vertices then slide along $Q$ to minimize the total length.  Varying a regular vertex $p$ along $Q$ imposes the condition
\begin{equation}
 \sum_{i=1}^{3}
 \frac{d\ell_i(p)}{ds_Q}=0,
 \label{eq:ee-vertex-balance}
\end{equation}
where $\ell_i(p)$ is the incident geodesic length in branch $i$ and $s_Q$ is proper distance along $Q$.

\subsection{Mutual and tripartite information 
}
\label{subsec:junction-information}

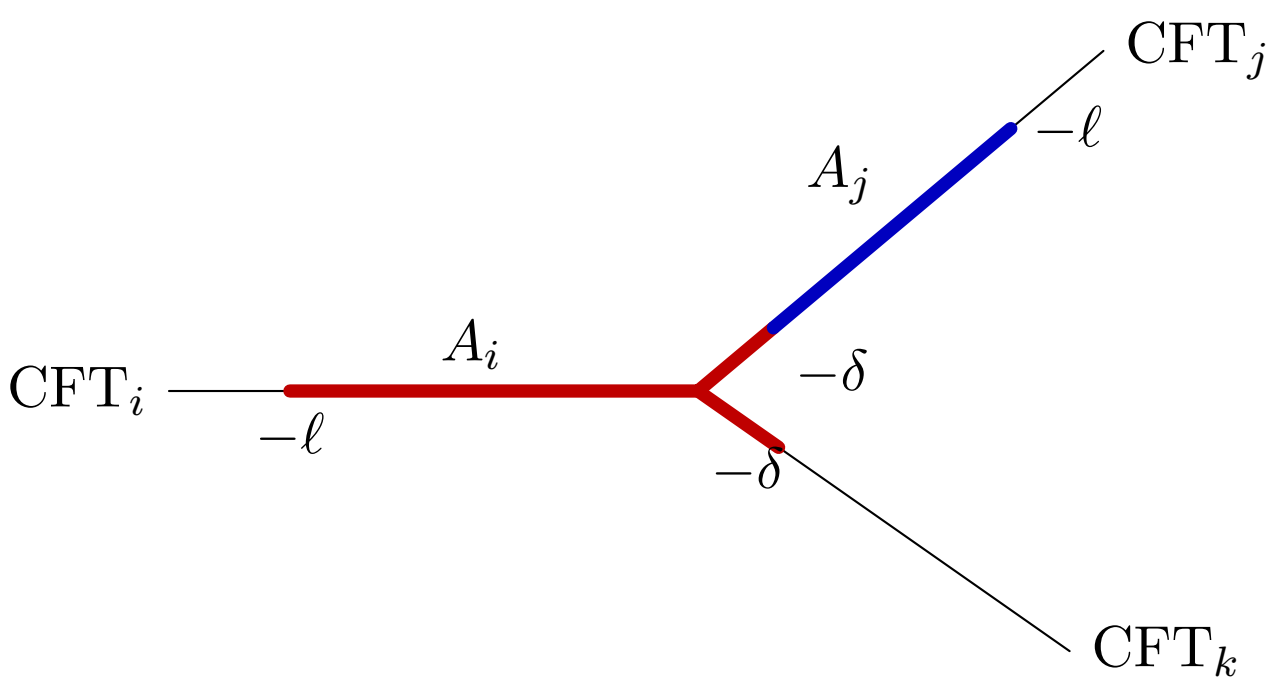
\begin{figure}[h!]
\centering
\begin{tikzpicture}[font=\small,line cap=round]
\draw[black,thin] (-3.5,0)--(0,0)--(40:3.5);\draw[black,thin] (0,0)--(-35:3);
\draw[red!75!black,line width=2.5pt] (-2.7,0)--(0,0)--(40:.65);\draw[red!75!black,line width=2.5pt] (0,0)--(-35:.65);
\draw[blue!75!black,line width=2.5pt] (40:.65)--(40:2.7);
\node[above] at (-1.5,0) {$A_i$};\node[above left] at (40:1.7) {$A_j$};
\node[below] at (-2.7,0) {$-\ell$};\node[right] at (40:2.7) {$-\ell$};
\node[below right] at (40:.65) {$-\delta$};\node[below] at (-35:.4) {$-\delta$};
\node[left] at (-3.5,0) {$\mathrm{CFT}_i$};\node[right] at (40:3.5) {$\mathrm{CFT}_j$};\node[right] at (-35:3) {$\mathrm{CFT}_k$};
\end{tikzpicture}
\caption{\small Regulated pair adjacent to the node in \eqref{eq:junction-i3-regions}. Red denotes $A_i$, including the node neighborhood; blue denotes $A_j$. Thin black lines show the complement.}
\label{fig:mutual-regions}
\end{figure}

We choose three disjoint boundary regions with a common macroscopic scale $\ell$ and assign the regulated neighborhood of the junction to one of them. 
For distinct $i,j,k$, we take 
\begin{equation}
 A_i=[-\ell,0]_i\cup(-\delta,0]_j\cup(-\delta,0]_k,
 \quad A_j=[-\ell,-\delta]_j,\quad A_k=[-\ell,-\delta]_k,
 \quad \delta=\lambda\epsilon .
 \label{eq:junction-i3-regions}
\end{equation}
Here $\ell>0$, $\epsilon$ is the UV cutoff and we hold $\lambda>1$ fixed as $\epsilon/\ell\to0$. This assignment keeps the regions disjoint while treating the node neighborhood consistently in every entropy combination. Fig.~\ref{fig:mutual-regions}  shows the regions $A_i,A_j$. We use  $S_i=S(A_i)$, $S_{ij}=S(A_i\cup A_j)$ and $S_{123}=S(A_1\cup A_2\cup A_3)$.

\paragraph{Mutual information.}
Subadditivity ensures that the pairwise mutual information is nonnegative, $I(A_i:A_j)=S_i+S_j-S_{ij}\geq 0$. For the regulated partition~\eqref{eq:junction-i3-regions}, the logarithmic contributions from the outer endpoints cancel in this combination. The remaining logarithm comes from the cut near the junction. 
Separating branch $r$ from the other two incurs a leading cost proportional either to $c_r$ or to $c_s+c_t$. 
The RT minimization selects
\begin{equation}
 h_r=\min(c_r,c_s+c_t),\qquad \{r,s,t\}=\{1,2,3\}.
 \label{eq:mi-single-cut-charge}
\end{equation}
As derived in App.~\ref{app:mutual-information}, the mutual information and its effective central charge are
\begin{align}
 I(A_i:A_j)&=\frac{c_{ij}^{\mathrm{eff}}}{3}\log\frac{\ell}{\epsilon}+\mathcal{O}(1),
 \label{eq:mi-leading}\\
c_{ij}^{\mathrm{eff}}&=3\lim_{\epsilon/\ell\to0}
 \frac{I(A_i:A_j)}{\log(\ell/\epsilon)}
 =\frac{h_i+h_j-h_k}{2}.
 \label{eq:mi-effective-charge}
\end{align}
The pairwise effective central charge $c_{ij}^{\mathrm{eff}}$
generalizes the interface effective central charge $c_{\mathrm{eff}}$~\cite{Afxonidis:2025jph, KarchEtAlUniversality2023}
to a selected pair of branches in the junction. When the central charges satisfy the triangle inequalities $c_i+c_j>c_k$ and its cyclic permutations,~ $c_{ij}^{\mathrm{eff}}=(c_i+c_j-c_k)/2$. 
In contrast, $c_{ij}^{\mathrm{eff}}=0$ if $c_k\geq c_i+c_j$, while $c_{ij}^{\mathrm{eff}}=c_j$ if $c_i\geq c_j+c_k$, with the analogous expression upon exchanging $i$ and $j$. Hence
\be
0\leq c_{ij}^{\mathrm{eff}}\leq\min(c_i,c_j).
\ee
A zero logarithmic coefficient does not exclude a finite mutual information. 

\paragraph{Tripartite information.}
The tripartite information compares the pairwise mutual information with its value conditioned on the third region~\cite{HaydenHeadrickMaloney2013,BalasubramanianEtAl2024}:
\begin{align}
\begin{split}
 I_3&=S_i+S_j+S_k-S_{ij}-S_{ik}-S_{jk}+S_{123}\\
    &=I(A_i:A_j)-I(A_i:A_j\mid A_k),
 \label{eq:junction-i3-information-identities}
 \end{split}
\end{align}
where $I(A_i:A_j\mid A_k)=S_{ik}+S_{jk}-S_k-S_{123}$. For a fixed choice of the three regions, $I_3$ is symmetric under their permutations. In Sec.~\ref{subsec:ee-inequalities}, we prove that the branched RT prescription obeys monogamy of mutual information (MMI), giving $I_3\leq0$. 

For the RT configurations analyzed in the two radius families of Sec.~\ref{sec:junction-transmission-bounds}, the logarithmic terms and the contributions associated with vertices near the node cancel in $I_3$. Their optimized lengths give the finite quantity $J=4G_NI_3$, 
\begin{equation}
 J=\begin{cases}
 E_{13}+E_{23}-E_3, & L_1\leq L_2<L_3,\quad L_1+L_2<L_3,\\[3pt]
 E_1+2E_2+a\log\dfrac{1-s_1}{2}+2b\log\dfrac{1-s}{2},
 & L_1=a\leq L_2=L_3=b.
 \end{cases}
 \label{eq:junction-i3-definition}
\end{equation}
Here $s_r=\sin\theta_r$ and $s=s_2=s_3$ in the second family. The optimized lengths $E_r$ and $E_{rs}$ are specified in App.~\ref{app:tripartite-information}. In limit $\epsilon\to 0$, $I_3=J/4G_N$ depends on the radii and tension through the background geometry and is independent of $\ell$, $\epsilon$ and the fixed ratio $\lambda$. Thus the correction used in~\eqref{eq:junction-corrected-transmission-bound} comes from the finite excess of conditional over ordinary mutual information.

\subsection{Entropy inequalities}
\label{subsec:ee-inequalities}
\label{subsec:ee-subadditivity}
\label{subsec:ee-monogamy}

We now establish subadditivity and monogamy of  mutual information (MMI) using contraction maps~\cite{BaoEtAl2015Cone,BaoNaskar2024}. 
We consider mutually disjoint boundary subregions $A,B,C$ and write $AB=A\cup B$, similarly for other unions. The argument relies on the common-trace condition~\eqref{eq:ee-common-trace} and the additivity of the classical RT length functional. 

Consider an inequality of the form $\sum_{a=1}^m S(X_a)\geq\sum_{b=1}^n S(Y_b)$. The minimizing homology regions $\gamma_{X_a}$ partition each bulk branch into cells with membership labels $x=(\chi_{r_{X_1}},\ldots,\chi_{r_{X_m}})\in\{0,1\}^m$. A map $f:\{0,1\}^m\to\{0,1\}^n$ defines trial regions $\widetilde \gamma_{Y_b}$ through the output bits. The map must satisfy two conditions~\cite{BaoNaskar2024}. 

First, $f$ must preserve the prescribed boundary anchoring. For each elementary boundary region $R$ (including the complement), define its input and output membership bits by
\begin{equation}
 x^{(R)}_a=\begin{cases}1,&R\subseteq X_a,\\0,&R\not\subseteq X_a,\end{cases}
 \qquad\qquad
 y^{(R)}_b=\begin{cases}1,&R\subseteq Y_b,\\0,&R\not\subseteq Y_b.\end{cases}
 \label{eq:ee-occurrence-bits}
\end{equation}
The boundary condition is
\begin{equation}
 f(x^{(R)})=y^{(R)}.
 \label{eq:ee-map-boundary}
\end{equation}
Physically, this ensures that each reconstructed bulk region is anchored on the prescribed boundary subsystem. Second, the map $f$ must not increase Hamming distance:
\begin{equation}
 d_H(f(x),f(y))\leq d_H(x,y).
 \label{eq:ee-hamming-contraction}
\end{equation}
For bitstrings $u,v\in\{0,1\}^k$, the Hamming distance is
\begin{equation}
 d_H(u,v)=\sum_{j=1}^k|u_j-v_j|.
 \label{eq:ee-hamming-definition}
\end{equation}
This prevents any RT segment from being counted more often in the output boundaries than in the input boundaries, so reassembly cannot increase the total length.

Each input region satisfies eq.~\eqref{eq:ee-common-trace}, so the input label on $Q$ is identical in all three branches. Applying the same map gives identical output labels there as well. 
The resulting trial regions therefore glue consistently across $Q$and satisfy the homology condition for the output regions.
We can thus apply the usual contraction-map argument to the
junction.

Divide the input RT prescription into segments $e$ separating cells labeled $x_e$ and $y_e$. A segment $e$ belongs to the boundary of an input region precisely when the corresponding membership bit differs across it. Its length is therefore counted once for each differing bit, giving a total multiplicity $d_H(x_e,y_e)$. The same argument gives $d_H(f(x_e),f(y_e))$ for the output boundaries. Consequently,
\begin{align}
 \sum_{b=1}^n\mathcal L(\widetilde\Gamma_{Y_b})
 &=\sum_e\ell(e)d_H(f(x_e),f(y_e))
 \leq\sum_e\ell(e)d_H(x_e,y_e)
 =\sum_{a=1}^m\mathcal L(\Gamma_{X_a}),
 \label{eq:ee-contraction-length}
\end{align}
where $\ell(e)$ is the proper length in the corresponding bulk branch. Junction vertices and intersections carry no length contribution. Coincident segments are counted with their multiplicities. Minimizing each output surface then gives
\begin{equation}
 \sum_{b=1}^n S(Y_b)
 \leq\frac{1}{4G_N}\sum_{b=1}^n\mathcal L(\widetilde\Gamma_{Y_b})
 \leq\sum_{a=1}^m S(X_a).
 \label{eq:ee-contraction-bound}
\end{equation}

\paragraph{Subadditivity.}
For inputs $(A,B)$ and output $AB$, we choose the union map 
\begin{equation}
 f(00)=0,\qquad f(10)=f(01)=f(11)=1.
 \label{eq:ee-subadditivity-map}
\end{equation}
The boundary labels on $A$, $B$, and their complement are $10$, $01$, and $00$, which map to the correct membership in $AB$. The map contracts Hamming distance: distinct inputs differ in at least one bit, while outputs differ in at most one. Its trial region is simply $\widetilde \gamma_{AB}=\gamma_A\cup \gamma_B$. Eq.~\eqref{eq:ee-contraction-bound} therefore yields
\begin{equation}
 S(A)+S(B)\geq S(AB),
 \qquad I(A:B)\equiv S(A)+S(B)-S(AB)\geq0.
 \label{eq:ee-subadditivity}
\end{equation}

For the reverse direction, the boundary condition would require $g(1)=10$ on $A$ and $g(1)=01$ on $B$. These requirements are incompatible, so no map satisfying the boundary condition exists.

\paragraph{Monogamy of mutual information.}
To map the inputs $(AB,AC,BC)$ to the outputs $(A,B,C,ABC)$, we choose 
\begin{equation}
 \begin{array}{c|c@{\qquad}c|c}
 x&f(x)&x&f(x)\\ \hline
 000&0000&100&0001\\
 001&0001&101&0101\\
 010&0001&110&1001\\
 011&0011&111&0001
 \end{array}.
 \label{eq:ee-contraction-map}
\end{equation}
The boundary labels on $A$, $B$, $C$, and their complement are $110$, $101$, $011$, and $000$. Their images $1001$, $0101$, $0011$, and $0000$ give precisely the required output regions. On each of the twelve edges of the input cube, the output labels differ by one bit. The triangle inequality along a shortest edge path then establishes Eq.~\eqref{eq:ee-hamming-contraction} for every pair of inputs. Applying Eq.~\eqref{eq:ee-contraction-bound} gives
\begin{equation}
 S(AB)+S(AC)+S(BC)\geq S(A)+S(B)+S(C)+S(ABC).
 \label{eq:ee-mmi}
\end{equation}
For the reverse direction, boundary anchoring requires $g(1001)=110$, $g(0101)=101$, $g(0011)=011$, and $g(0000)=000$. The input $0001$ is distance one from all four boundary labels. Contraction would therefore require its image to be within distance one of $110$, $101$, $011$, and $000$ simultaneously. The last condition restricts the image to $000,100,010,001$, and each fails at least one of the other conditions. Thus no reverse map satisfies both requirements. This rules out a reverse contraction proof; it does not imply strictness in every configuration.

Equivalently, the tripartite information obeys
\begin{equation}
\begin{aligned}
I_3(A:B:C)
&\equiv S(A)+S(B)+S(C) \\
&\quad -S(AB)-S(AC)-S(BC)+S(ABC)
\leq 0 .
\end{aligned}
\label{eq:ee-tripartite-information}
\end{equation}
This is also equivalent to $I(A:BC)\geq I(A:B)+I(A:C)$. Unlike subadditivity, MMI is not a general quantum entropy inequality; it is a constraint on classical RT entropies~\cite{HaydenHeadrickMaloney2013}.

The arguments do not depend on which branches contain the boundary regions or on the topology of the minimizing RT configuration. 
Their validity rests on common-trace homology and additive classical lengths; we do not include bulk quantum entropy or brane-localized entropy contributions. 

\section{Bounds on multipartite energy transmission}
\label{sec:junction-transmission-bounds}

For a two-CFT interface, the proposed bound $c_{LR}\leq c_{\mathrm{eff}}\leq\min(c_L,c_R)$  related energy transmission to interface entanglement~\cite{Karch}. A Y-junction instead  distributes transmitted energy among two outgoing channels, while the third leg also influences pairwise correlations. For the holographic junction, grouping the other two branches into a single subsystem gives the one-versus-rest bound on the total transmitted flux~\cite{LiuWangJunction2025}
\begin{equation}
 c_i\sum_{j\ne i}\mathcal T_{ij}\leq c_{\mathrm{eff}}^{i}=\min(c_i,c_j+c_k),
 \qquad \{i,j,k\}=\{1,2,3\}.
 \label{eq:junction-one-versus-rest-bound}
\end{equation}
Here $c_{\mathrm{eff}}^{i}=\min(c_i,c_j+c_k)$ is the one-versus-rest effective central charge~\cite{LiuWangJunction2025}. 
We now compare individual transmission channels with the pairwise information coefficients obtained from the branched RT prescription in Sec.~\ref{subsec:rt-networks}. We first identify violations of the direct pairwise bound and then show how the finite tripartite information restores it for the regulated equal-macroscopic-length regions in two families of junction geometries.

\subsection{Failure of the direct pairwise interface bound}
\label{subsec:junction-pairwise-failure}

For the regulated regions with common macroscopic length defined in~\eqref{eq:junction-i3-regions}, we compare the weighted transmission coefficient $c_{ij}$ with $c_{ij}^{\mathrm{eff}}$: 
\begin{equation}
 c_{ij}=c_i\mathcal T_{ij}
 =\frac{3L_a}{G_N}\frac{\eta_i\eta_j}{E},\qquad
 c_{ij}^{\mathrm{eff}}=\frac{h_i+h_j-h_k}{2},
 \qquad\{i,j,k\}=\{1,2,3\}.
 \label{eq:junction-energy-transmission}
\end{equation}
Here the transport weights $\eta_i$, their sum $E$ and the one-versus-rest coefficients $h_i$  are defined in Secs~\ref{subsec:hj-transport} and~\ref{subsec:junction-information}. We also  write $s_i=\sin\theta_i$. The physical transport coefficients satisfy $0\leq c_{ij}\leq\min(c_i,c_j)$. For the two radius families, Eq.~\eqref{eq:mi-effective-charge} yields 
\begin{equation}
 \begin{array}{c@{\quad}|@{\quad}c@{\quad}c@{\quad}c}
&c^{\mathrm{eff}}_{12}&c^{\mathrm{eff}}_{13}&c^{\mathrm{eff}}_{23}\\[2pt] \hline
 L_1+L_2<L_3&0&c_1&c_2\\[3pt]
 L_1\leq L_2=L_3&c_1/2&c_1/2&c_2-c_1/2
 \end{array}
 \label{eq:junction-effective-charge-regimes}
\end{equation}
Both families obey $0\leq c_{ij}^{\mathrm{eff}}\leq\min(c_i,c_j)$, but a direct pairwise extension of the interface bound would further require $c_{ij}\leq c_{ij}^{\mathrm{eff}}$.

\paragraph{The dominant-radius family: $L_1\leq L_2<L_3$ with $L_1+L_2<L_3$.}
We consider the tension range~\cite{LiuWangJunction2025}
\begin{equation}
 \frac1{L_1}+\frac1{L_2}-\frac1{L_3}
 <\tau<\frac1{L_1}+\frac1{L_2}+\frac1{L_3},\qquad
 \tau=\sum_{i=1}^3\frac{s_i}{L_i}.
 \label{eq:transport-dominant-tension}
\end{equation}
We parametrize the two geometric branches by 
\begin{equation}
 s_1=\sqrt{1-L_1^2/L_a^2},\quad
 s_2=\sqrt{1-L_2^2/L_a^2},\quad
 s_3=\sigma\sqrt{1-L_3^2/L_a^2},\quad
 L_a\geq L_3,\quad\sigma=\pm1.
 \label{eq:transport-dominant-branches}
\end{equation}
The branches meet at $L_a=L_3$. For every finite $L_a$,~\eqref{eq:junction-energy-transmission} gives $\mathcal T_{12}>0$, while~\eqref{eq:junction-effective-charge-regimes} gives $c^{\mathrm{eff}}_{12}=0$. Consequently,
\begin{equation}
 c_1\mathcal T_{12}>c^{\mathrm{eff}}_{12}=0.
 \label{eq:transport-dominant-violation}
\end{equation}
The other two pairs satisfy the bound throughout this range:
\begin{equation}
 0\leq c_{13}\leq c_1=c_{13}^{\mathrm{eff}},\qquad
 0\leq c_{23}\leq c_2=c_{23}^{\mathrm{eff}}.
 \label{eq:transport-dominant-valid-channels}
\end{equation}
These inequalities follow directly from the transport bound $c_{ij}\leq\min(c_i,c_j)$. Thus only the channel between the two smaller branches violates the pairwise bound. Its positive transmitted flux coexists with a vanishing logarithmic coefficient of mutual information. This does not require the full mutual information to vanish.

\paragraph{The equal-larger-radius family: $L_1=a\leq L_2=L_3=b$.}
We take $0<a<b$, set $\rho=a/b\in(0,1)$, and follow the physical branch with $s_1>0$, $s_2=s_3=s>0$, and $1/a\leq\tau<1/a+2/b$. At its lower tension endpoint, define
\begin{equation}
 s_0=\frac{\rho}{2+\sqrt{4-3\rho^2}},\qquad
 L_{a,\min}=\frac{b}{\sqrt{1-s_0^2}},\qquad
 s\geq s_0,\quad L_a\geq L_{a,\min}.
 \label{eq:transport-equal-minimum}
\end{equation}
At $\tau=1/a$, one has $s=s_0$ and $s_1=1-2\rho s_0$. Substitution into the transport coefficient gives
\begin{equation}
 \left.\mathcal T_{23}\right|_{\tau=1/a}
 =\frac{1-s_0}{1+s_0},\qquad
 \left.\mathcal T_{23}\right|_{\tau=1/a}
 -\frac{c^{\mathrm{eff}}_{23}}{c_2}
 =\frac{2s_0^2(1-3s_0)}{(1+s_0)(1+3s_0^2)}>0.
 \label{eq:transport-equal-violation}
\end{equation}
The strict sign follows from $0<s_0<1/3$. Thus every fixed ratio $0<a/b<1$ has a range near the lower tension endpoint where the direct pairwise bound fails. For example, $a=1$, $b=3$ gives $L_{a,\min}\simeq3.010934$ and $\mathcal T_{23}\simeq0.843070$, exceeding $c^{\mathrm{eff}}_{23}/c_2=5/6$.

The pairs involving the smaller branch satisfy the bound throughout the tension range. Since $c_{12}^{\mathrm{eff}}=c_{13}^{\mathrm{eff}}=c_1/2$, it suffices to show $\mathcal T_{12}=\mathcal T_{13}<1/2$:
\begin{equation}
 \mathcal T_{12}=\frac{2b}{a(1+s)+2b(1+s_1)},\qquad
 \left.\mathcal T_{12}\right|_{\tau=1/a}
 =\frac{1+3s_0^2}{2(1+s_0)}<\frac12,
 \label{eq:transport-equal-small-channel}
\end{equation}
The denominator increases with $L_a$, so the endpoint value is the maximum. Hence
\begin{equation}
 c_{12}<c_{12}^{\mathrm{eff}},\qquad
 c_{13}<c_{13}^{\mathrm{eff}}\qquad(0<a<b).
 \label{eq:transport-equal-valid-channels}
\end{equation}
For the equal-radius endpoint $a=b$, $s_1=s_2=s_3=s\geq1/3$ in the stated tension range. All three channels then obey $\mathcal T_{ij}=2/[3(1+s)]\leq1/2$, with equality at the lower endpoint.

Unlike the dominant-radius example, this violation occurs even though $c_{23}^{\mathrm{eff}}>0$. 
At fixed radii, the effective charges are tension-independent, whereas the transmitted fluxes vary with tension. In the equal-larger-radius family, $\mathcal T_{23}$ decreases to $b/(a+2b)<1-a/(2b)$ at the upper tension endpoint, so the bound is restored at sufficiently large tension.

Together, these examples show that $c_{ij}^{\mathrm{eff}}$ does not universally bound an individual transmission channel of a Y-junction, even though some channels obey the pairwise inequality. 
This is a failure of the direct pairwise extension to a three-branch junction, where the remaining branch stays coupled to the selected pair.

A channel can therefore exceed its pairwise effective charge while the total flux remains bounded by the one-versus-rest charge:
\begin{equation}
 c_i\sum_{j\ne i}\mathcal T_{ij}
 \leq c_{\mathrm{eff}}^{i}
 =\min(c_i,c_j+c_k),\qquad\{i,j,k\}=\{1,2,3\}.
 \label{eq:junction-recovered-total-bound}
\end{equation}
Grouping the other two branches into one subsystem gives an effective bipartition $i\mid(jk)$. The total flux then obeys the same energy-versus-information bound as a two-CFT interface, with $c_j+c_k$ as the central charge of the combined side. Thus the pairwise violations coexist with the interface-type bound for the total channel~\cite{LiuWangJunction2025}.

\subsection{A tripartite correction to the transmission bound}
\label{subsec:junction-tripartite-bound}

The violations in Sec.~\ref{subsec:junction-pairwise-failure} show that the pairwise effective central charges alone do not bound every energy-transmission channel of a junction. The third subsystem remains coupled to the selected pair, so information involving all three subsystems must also be considered. We use the finite tripartite information summarized in~\eqref{eq:junction-i3-definition} and derived in App.~\ref{app:tripartite-information}. Its sign is fixed by MMI, and $-I_3=I(A_i:A_j\mid A_k)-I(A_i:A_j)$ measures the excess of conditional over ordinary mutual information. We will prove that, in the two radius families considered here, it restores the channel bound: 
\begin{equation}
 \quad c_i\mathcal T_{ij}\leq c_{ij}^{\mathrm{eff}}+3(-I_3)
 \leq\min\{c_i+3(-I_3),c_j+3(-I_3)\}.\quad
 \label{eq:junction-corrected-transmission-bound}
\end{equation}
The second inequality follows from the pairwise effective-charge bound. The first is proved below using the finite expressions for $J=4G_NI_3$. The coefficient $3$ is fixed by limiting saturation, rather than by MMI alone. To compare the curves, retain
\begin{equation}
 \mathcal R_{ij}=\frac{c_i\mathcal T_{ij}}{c_{ij}^{\mathrm{eff}}+3(-I_3)}.
 \label{eq:transport-plot-ratio}
\end{equation}
The corrected channel bound is $\mathcal R_{ij}\leq1$.

\paragraph{The dominant-radius family: $L_1\leq L_2<L_3$ with $L_1+L_2<L_3$.} 
We need only control the $12$ channel, because $c_{13}^{\mathrm{eff}}=c_1$ and $c_{23}^{\mathrm{eff}}=c_2$ already bound the other two fluxes. Setting $M=L_1+L_2$ and using $\phi_i$ defined in~\eqref{eq:transport-phi}, we introduce the auxiliary maximum
\begin{equation}
 \mathcal E(p)=\max_{t\geq0}\{pt-L_3\phi_3(t)\},\qquad0\leq p\leq M.
\end{equation}
Because $\phi_1,\phi_2\geq0$, Eq.~\eqref{eq:transport-dominant-variational} gives $E_{13}\leq\mathcal E(L_2)$, $E_{23}\leq\mathcal E(L_1)$, and $E_3=\mathcal E(M)$. The maximizer $t_p$ satisfies $L_3\phi_3'(t_p)=p$. Since $0<p\leq M<L_3$, it lies on the branch where $\phi_3''(t_p)>0$; moreover, $\phi_3''(t_p)\leq\phi_3''(0)=1/(1+s_3)$. Thus
\begin{equation}
 \mathcal E''(p)=\frac1{L_3\phi_3''(t_p)}\geq\frac{1+s_3}{L_3},
 \qquad\phi_3''(t)=\frac{1+s_3\cosh t}{(\cosh t+s_3)^2}.
 \label{eq:transport-curvature-bound}
\end{equation}
This holds on both geometric branches $\sigma=\pm1$. Integrating the curvature bound and using $\mathcal E(0)=0$, we find 
\begin{align}
 -J&\geq\mathcal E(M)-\mathcal E(L_1)-\mathcal E(L_2)\nonumber\\
 &=\int_0^{L_1}\!du\int_0^{L_2}\!dv\,\mathcal E''(u+v)
 \geq\frac{L_1L_2}{L_3}(1+s_3).
 \label{eq:transport-dominant-i3-bound}
\end{align}
The transport weights satisfy $\eta_3(\eta_1+\eta_2)<1$ in this radius range, implying $\eta_3(1+\eta_1^2)(1+\eta_2^2)<E$. Hence
\begin{equation}
 \frac{L_1L_2}{L_3}(1+s_3)
 =\frac{4L_a\eta_1\eta_2}{\eta_3(1+\eta_1^2)(1+\eta_2^2)}
 >\frac{4L_a\eta_1\eta_2}{E}=2L_1\mathcal T_{12}.
 \label{eq:transport-dominant-comparison}
\end{equation}
It follows that $3(-I_3)>c_1\mathcal T_{12}$ at finite $L_a$. Thus all three corrected channel bounds hold on both branches for any $L_i$ within this family. 

\begin{figure}[htbp]
 \centering\includegraphics[width=0.95\textwidth]{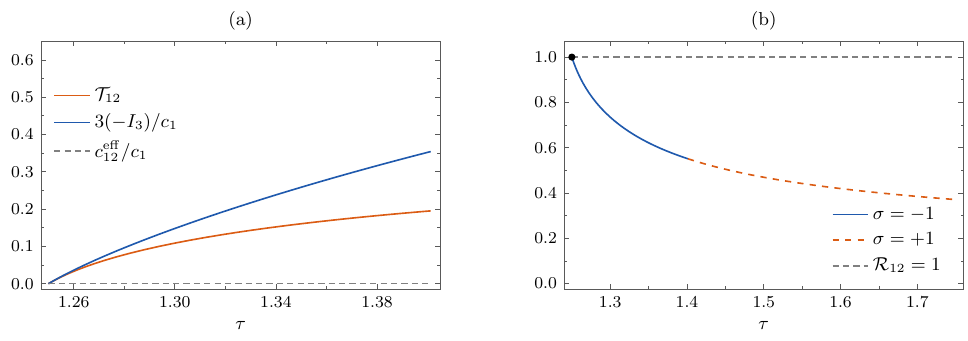}
 \vspace{-0.4cm}
 \caption{\small The dominant-radius family with $L_1:L_2:L_3=1:2:4$. (a) Transmitted flux and corrected bound for $\sigma=-1$. (b) Corrected ratio on both branches; the point at $\tau=5/4$ denotes the saturation limit.}
 \label{fig:junction-dominant-bounds}
\end{figure}

As a explicit numerical example with a special $L_1:L_2:L_3=1:2:4$, Fig.~\ref{fig:junction-dominant-bounds}(a) shows how the tripartite term bounds the transmitted flux even though $c_{12}^{\mathrm{eff}}=0$, while panel (b) shows $\mathcal R_{12}<1$ across the tension range.

\paragraph{The equal-larger-radius family: $L_1=a\leq L_2=L_3=b$.}
The $12$ and $13$ channels already satisfy their pairwise bounds, so the nonnegative correction preserves them. For the $23$ channel, use the quantities $E_1,E_2$ in~\eqref{eq:transport-equal-variational}. From $\phi_b(t)\geq2\log\cosh(t/2)$ and $\phi_a(t)\geq0$,
\begin{align}
\begin{split}
 E_1&\leq(2b+a)\log\left(1+\frac{a}{2b}\right)
 +(2b-a)\log\left(1-\frac{a}{2b}\right)
 \leq a\rho\log\frac{27}{16},\\
 E_2&\leq b\log[2(1+s)],\qquad\rho=a/b.
 \label{eq:transport-equal-energy-estimates}
\end{split}
\end{align}
The last bound on $E_1$ follows from its positive power series for $0<\rho\leq1$. Substitution into~\eqref{eq:transport-equal-i3}, followed by $s_1\geq1/3$, $s\geq s_0\geq\rho/4$ and $-\log(1-s^2)\geq s^2$, yields
\begin{equation}
 -J\geq a\left[\log3-\rho\log\frac{27}{16}+\frac{\rho}{8}\right]
 \geq a\left[\log\frac{16}{9}+\frac18\right]>\frac{3a}{5}.
 \label{eq:transport-equal-uniform-gap}
\end{equation}
The excess of transmitted flux over the pairwise bound satisfies
\begin{equation}
 2b\mathcal T_{23}-(2b-a)
 \leq a\frac{2+\rho}{4+\rho}\leq\frac{3a}{5},
 \label{eq:transport-equal-excess-bound}
\end{equation}
where the first inequality follows from $\mathcal T_{23}\leq4b/(4b+a)$. Comparing these two estimates proves the corrected bound for the $23$ channel as well. 

\begin{figure}[h!]
\centering\includegraphics[width=0.95\textwidth]{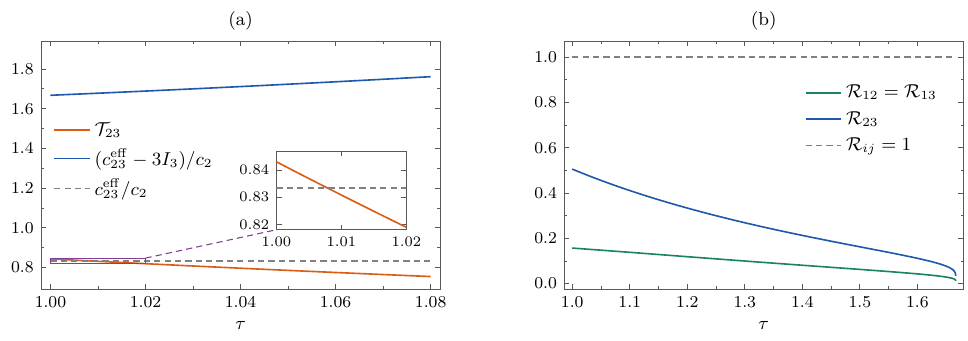}
\vspace{-0.1cm}
 \caption{\small The equal-larger-radius family with $L_1=1$, $L_2=L_3=3$. (a) Original and corrected bounds; inset: low-tension pairwise violation. (b) Corrected channel ratios.}
 \label{fig:junction-triangle-bounds}
\end{figure}

An illustration figure is shown in Fig.~\ref{fig:junction-triangle-bounds}(a), the inset resolves the low-tension violation of the pairwise bound, while the corrected curve remains above the transmitted flux. Panel (b) shows that the corrected ratios stay below one for every pair.

\paragraph{Saturation of the corrected bound.}
We first examine the lower-tension limit of the dominant-radius family on the $\sigma=-1$ branch. With $\tau_-=L_1^{-1}+L_2^{-1}-L_3^{-1}$, expanding the transmission coefficient and the independently minimized RT lengths at large $L_a$ gives 
\begin{equation}
 \mathcal T_{12}=\frac{L_2L_3}{4L_a^2}+\mathcal{O}(L_a^{-4}),\quad
 -J=\frac{L_1L_2L_3}{2L_a^2}+\mathcal{O}(L_a^{-4}),\quad
 \tau-\tau_-=\frac{L_3-M}{2L_a^2}+\mathcal{O}(L_a^{-4}).
 \label{eq:transport-sharp-saturation}
\end{equation}
Expanding the vertex equations of App.~\ref{app:tripartite-information} to the next order, we obtain\begin{equation}
 \mathcal R_{12}=1-\frac{3[L_1L_2+ML_3]}{8L_a^2}+\mathcal{O}(L_a^{-4})
 \rightarrow1\qquad(\tau\to\tau_-^+).
 \label{eq:transport-saturation-approach}
\end{equation}
The ratio approaches one from below, as in Fig.~\ref{fig:junction-dominant-bounds}(b). Both $c_1\mathcal T_{12}$ and $3(-I_3)$ tend to zero with the same leading coefficient. Equality is therefore a limiting saturation, not an equality at finite $L_a$. Any smaller coefficient multiplying $-I_3$ would violate the bound near this endpoint, so $3$ is sharp. The upper inequality in~\eqref{eq:junction-corrected-transmission-bound} is separately saturated by the $13$ and $23$ channels throughout this family.

For the equal-larger-radius family,~\eqref{eq:transport-equal-uniform-gap} and~\eqref{eq:transport-equal-excess-bound} instead imply a uniform positive gap:
\begin{equation}
 c_{23}^{\mathrm{eff}}+3(-I_3)-c_2\mathcal T_{23}
 \geq\frac{3a}{4G_N}\left[\log\frac{16}{9}+\frac18-\frac35\right]>0.
 \label{eq:transport-equal-strict-gap}
\end{equation}
The bracket is approximately $0.100364$. For the other two channels, their original pairwise bounds give a corrected gap of at least $3(-I_3)>0$. These estimates also hold at $a=b$ on the positive-angle branch. Thus the corrected lower bound cannot be saturated in this family. In particular, at the upper tension endpoint the finite formula gives $-J=2(a+2b)\log(L_a/b)+\mathcal{O}(1)$ while the transmitted flux remains bounded, so all $\mathcal R_{ij}\to0$. This contrasts with the dominant-radius saturation limit and agrees with Fig.~\ref{fig:junction-triangle-bounds}(b).

\section{Entropic transmission bounds in free-field junctions}
\label{sec:free-field-bounds}
We now test the entropic transmission bounds beyond holography in junctions of a real free boson and a complex Dirac fermion, both with $c=1$. We first construct analytic counterexamples to the pairwise bound and examine whether tripartite information repairs them. We then use Gaussian canonical purification to test a correction based on L- entropy, carefully distinguishing finite intervals from complete-leg partitions. The corresponding Majorana results follow by halving the Dirac central-charge-weighted quantities. 

We first summarize the useful analytic interface functions and their normalization
in Y-junctions with free fields. 
Let $s\in[0,1]$ denote the transmission amplitude, so that $\mathcal T=s^2$. 
We use the known interface effective central charges $f_{\rm B}(s)$ and $f_{\rm D}(s)$ for a real boson and a complex Dirac fermion~\cite{Karch,GM,CMC}: 
\begin{align}
\begin{split}
 H(s)={}&(s+1)\log(s+1)\log s
 +(s-1)\operatorname{Li}_2(1-s)+(s+1)\operatorname{Li}_2(-s),\\
 f_{\rm B}(s)={}&\frac12+s+\frac{3}{\pi^2}H(s),\\
 f_{\rm D}(s)={}&s-1-\frac{6}{\pi^2}H(s).
 \label{eq:dirclosed}
 \end{split}
\end{align}
We define the endpoints by continuity, $f_a(0)=0$ and $f_a(1)=1$. A Majorana fermion contributes half the Dirac result, $f_{\rm M}=f_{\rm D}/2$, consistent with their central charges $1/2$ and $1$, respectively. To work directly with the transmission probability $t=s^2$, we set $F_a(t)=f_a(\sqrt t)$ and use the equivalent spectral integrals 
\begin{align}
\begin{split}
 F_{\rm D}(t)&=\frac{12}{\pi^2}\int_0^\infty dx\,\,h\!\left(\frac{1-\sqrt{1-t\operatorname{sech}^2x}}2\right),\\
 F_{\rm B}(t)&=\frac6{\pi^2}\int_0^\infty dx\,\,g\!\left(\frac12\sqrt{1+t\operatorname{csch}^2x}\right),\label{eq:bosint}\\
 h(p)&=-p\log p-(1-p)\log(1-p),\\
 g(\nu)&=(\nu+\tfrac12)\log(\nu+\tfrac12)-(\nu-\tfrac12)\log(\nu-\tfrac12).
 \end{split}
\end{align}
Here $h(p)$ and $g(\nu)$ are the single-mode fermionic and bosonic entropies, and $x$ parametrizes the entanglement spectrum. These integrals reproduce the dilogarithmic expressions and obey the two-leg interface bound $F_a(t)\geq t$ for $c=1$. At weak transmission,
 \begin{equation}
 F_{\rm B}(t)=\frac32\sqrt t+\mathcal{O}(t|\log t|),\qquad
 F_{\rm D}(t)=\frac{3t}{\pi^2}(3-\log t)+\mathcal{O}(t^2|\log t|).
 \label{eq:smalltransmission}
\end{equation}
The bosonic coefficient therefore scales as $\sqrt t$, whereas the Dirac coefficient scales as $t|\log t|$. We will use this difference to establish pairwise-bound violations analytically as the third leg decouples. 

\subsection{Counterexamples in free-field theories}
We couple three identical branches through a real symmetric orthogonal matrix $U$. It defines one-particle scattering for the Dirac fermions~\cite{CMC} and conformal current gluing for the real bosons~\cite{GM}. We regulate the bosonic zero mode; compactification-dependent constants and subleading zero-mode terms do not affect the leading coefficient of $W=\log(\ell/\epsilon)$ used below~\cite{GM}. 
The transmission probabilities are $\mathcal T_{ij}=U_{ji}^2$ and $t_i=1-U_{ii}^2$. 

We use the same equal-length boundary regions and node assignment as in~\eqref{eq:junction-i3-regions}. In coordinates increasing away from the node,
\begin{equation}
 A_1=(0,\ell)_1\cup(0,\delta)_2\cup(0,\delta)_3,\qquad
 A_2=(\delta,\ell)_2,\qquad A_3=(\delta,\ell)_3,
 \quad\delta=\lambda\epsilon,\quad\lambda>1.
 \label{eq:interval-ft}
\end{equation}
The three parties are disjoint, and the exterior of their union is traced out. The node neighborhood belongs only to $A_1$; the UV limit keeps $\lambda$ fixed, rather than $\delta/\ell$.
For the interval considered here, 
the leading entropy of $A_X=\bigcup_{i\in X}A_i$ is
\begin{equation}
 S_a(A_X)=\frac{|X|+\sum_\alpha F_a(t_\alpha)}6W+o(W),
 \qquad W=\log(\ell/\epsilon)
\end{equation}
here the index $a=\mathrm{B},\mathrm{D},\mathrm{M}$ labels the real boson, Dirac fermion, and Majorana, $t_\alpha$ are the eigenvalues of $1-U_{XX}^{T}U_{XX}$,\footnote{Here $U_{XX}$ denotes the restriction of the scattering matrix to the branches in $X$. By orthogonality, $I_{|X|}-U_{XX}^TU_{XX}=U_{\bar X X}^TU_{\bar X X}$, whose eigenvalues $t_\alpha$ are the transmission eigenvalues from $X$ to its complement. } and the term $|X|/6$ comes from the outer endpoints. 
Taking a two-branch subset as an example, the transmission eigenvalues are $\{0,t_k\}$; the outer-endpoint terms then cancel in the mutual information, giving
 \begin{equation}
 c^{\rm eff}_{ij,a}=3\lim_{W\to\infty}\frac{I_a(A_i:A_j)}W
 =\frac{F_a(t_i)+F_a(t_j)-F_a(t_k)}2.
 \label{eq:pairfield}
\end{equation}

To construct an analytically tractable counterexample, we consider a rank-one projector within the projector representation of scale-invariant junction matrices~\cite{BellazziniMintchevSorba2007}. Writing $P=vv^T$ with $v^Tv=1$, the resulting scattering matrix $U=I-2vv^T$ is a Householder reflection~\cite{Householder1958}. To preserve the symmetry between the first two legs, we choose 
\begin{equation}
 v=\left(\sqrt{(1-u)/2},~~\sqrt{(1-u)/2},~~\sqrt u\right)^T.
\end{equation}
Then $\mathcal T_{12}=(1-u)^2$, $t_1=t_2=1-u^2$, $t_3=4u(1-u)$ and
\begin{equation}
 c^{\rm eff}_{12,a}=F_a(1-u^2)-\tfrac12F_a(4u(1-u)).
\end{equation}
At $u=1/10$,
\begin{equation}
 U=\frac1{10}\begin{pmatrix}1&-9&-3\sqrt2\\-9&1&-3\sqrt2\\-3\sqrt2&-3\sqrt2&8\end{pmatrix}.
\end{equation}
Evaluating the dilogarithm formulas yields
\begin{center}
\begin{tabular}{c|ccc}
Theory & $c\mathcal T_{12}$ & $c^{\rm eff}_{12}$ & $D_{12}$\\\hline
Real boson & $0.810000$ & $0.660353$ & $0.149647$\\
Dirac fermion & $0.810000$ & $0.764256$ & $0.045744$\\
Majorana fermion & $0.405000$ & $0.382128$ & $0.022872$
\end{tabular}
\end{center}
Here we define $D_{ij}\equiv[c\mathcal T_{ij}-c^{\rm eff}_{ij}]_+$. Thus the direct pairwise extension fails in both statistics. This does not contradict the original two-CFT interface inequality.
The failure is analytic rather than an isolated numerical observation. Eq.~\eqref{eq:smalltransmission} implies
\begin{align}
\begin{split}
 D_{12,\rm B}&=\frac32\sqrt u+\mathcal{O}(u|\log u|),\\
 D_{12,\rm D}&=u\left[\frac6{\pi^2}(3-\log(4u))-2\right]+\mathcal{O}(u^2|\log u|),
\end{split}
\end{align}
which are positive for sufficiently small $u>0$.

\subsection{Tripartite information does not restore the bound}
 The holographic correction would read
\begin{equation}
 c\mathcal T_{ij}\leq c^{\rm eff}_{ij,a}+\kappa(-J_a),\qquad
 J_a=\lim_{\epsilon/\ell\to0}I_{3,a},\qquad\kappa\geq0,
 \label{eq:tripartite-correction-free}
\end{equation}
when the finite limit exists. This convention retains the finite part, with the holographic length normalization absorbed into $\kappa$.

For both bosons and Dirac fermions, the entropies have the same logarithmic structure, with the dependence on statistics encoded in the functions $F_a(t)$: 
\begin{align}
\begin{split}
S_i&=\frac{1+F_a(t_i)}{6}W+o(W),\\
S_{ij}&=\frac{2+F_a(t_k)}{6}W+o(W),\\
S_{123}&=\frac12 W+o(W),
\qquad W=\log(\ell/\epsilon).
\end{split}
\end{align}
The terms proportional to $W$ cancel in the tripartite information for both theories, irrespective of the explicit form of $F_a(t)$. We therefore examine the remaining contributions separately.


For the boson, the finite remainder need not cancel. We evaluate the seven Gaussian entropies from the symplectic eigenvalues of their restricted covariances, using one Neumann and two Dirichlet gluing channels, $U=2vv^T-1$, and massless noncompact fields with Dirichlet ends at $L=3\ell$. For $u=0.1$ and $\lambda=2$, increasing $n=\ell/\epsilon$ gives
\begin{equation}
\begin{array}{c|cccc}
 n&96&192&384&768\\\hline
 I_{3,\rm B}&0.063014&0.067728&0.070947&0.073147
\end{array}
\end{equation}
The data approach a positive finite remainder, with an extrapolated value around $J_{\rm B}\simeq0.078$ for  $n\gg 1$. Fits of the mutual information in this same partition remain consistent with $c^{\rm eff}_{12,\rm B}\simeq 0.66035$, leaving $D_{12,\rm B}\simeq 0.14965>0$. Hence the obstruction is its sign:
\begin{equation}
 D_{12,\rm B}>0,\qquad \kappa(-J_{\rm B})\leq0.
\end{equation}
No nonnegative $\kappa$ repairs this example in~\eqref{eq:tripartite-correction-free}. Unlike classical holographic entropy, free-field entropy is not constrained by holographic monogamy to have $I_3\leq0$.

Reversing the correction to $+\kappa J_{\rm B}$ defines a different candidate. It can repair this particular bosonic example, requiring approximately $\kappa\gtrsim1.9$ in the stated infrared geometry. The finite remainder depends on the infrared boundary conditions and zero-mode prescription.

For identical free Dirac fields, the continuum Gaussian calculation cancels the finite functions as well. Writing $d=\bigcup_i(0,\delta)_i$ and $F_i=(\delta,\ell)_i$, the regions are $(dF_1,F_2,F_3)$. Channel singular-value decomposition gives $S(F_i)=b_i$ and $S(F_iF_j)=b_*+b_k$; filling $d$ changes $S(F_X)$ by a function affine in $|X|$. Both contributions cancel in the seven-term combination. At fixed $\delta/\epsilon$, finite lattice corrections need not vanish identically, but the half-filled Gaussian chain gives $I_{3,\rm D}\to0$. For $u=0.1$ and $\lambda=2$, increasing $n=\ell/\epsilon$ yields
\begin{equation}
\begin{array}{c|cccc}
 n&48&96&192&384\\\hline
 I_{3,\rm D}&~~-2.286\times10^{-4}&~~-1.227\times10^{-4}&~~-6.362\times10^{-5}&~~-3.239\times10^{-5}
\end{array}
\end{equation}
Thus $J_{\rm D}=0$ for this realization, while the transmission deficit remains positive. The Majorana entropy coefficients and deficit are half the Dirac values and show the same obstruction.

The bosonic sign and the vanishing Dirac limit therefore prevent a single nonnegative $-I_3$ correction from repairing both free-field examples. 

\subsection{A candidate L-entropy correction}

We therefore examine L-entropy, a measure of genuine tripartite correlations that can remain nonzero for pure three-party states~\cite{LE}. We partition the pure junction ground state into its three complete physical legs $\mathcal H_i$. In contrast, the finite intervals used to define $c^{\rm eff}_{ij}$ have a mixed reduced state after tracing out their exterior. Let $S_R(i:j)$ be the reflected entropy of the two-leg reduced state, evaluated through its canonical purification. We define~\cite{DF,LE}
 \begin{equation}
 \ell_{ij}=2\min\{S(\mathcal H_i),S(\mathcal H_j)\}-S_R(i:j),\qquad
 \mathcal L_{123}=(\ell_{12}\ell_{13}\ell_{23})^{1/3}.
\end{equation}
The reflected-entropy upper bound~\cite{DF} makes each $\ell_{ij}$ nonnegative. Writing $W_L=\log(L/\epsilon)$, set
\begin{equation}
 S(\mathcal H_i)=\gamma_iW_L+o(W_L),\quad
 S_R(i:j)=r_{ij}W_L+o(W_L),\quad
 \beta_{ij}=2\min(\gamma_i,\gamma_j)-r_{ij}.
 \label{eq:rij}
\end{equation}
The coefficient entering the transmission bound is
\begin{equation}
 b=\lim_{W_L\to\infty}\frac{\mathcal L_{123}}{W_L}
   =(\beta_{12}\beta_{13}\beta_{23})^{1/3},\qquad
 \gamma_i=\frac{F_a(t_i)}6.
\end{equation}
Thus $b$ measures correlations among complete legs, while $c^{\rm eff}_{ij}$ measures the leading mutual information of the regulated equal-length regions. Their relation uses the endpoint separation discussed above; no finite-interval L-entropy is substituted for $b$.

The same candidate correction applies to the real boson and the Dirac fermion:
\begin{equation}
 c\mathcal T_{ij}\leq c^{\rm eff}_{ij,a}+\sqrt{cb_a}
 \qquad\text{equivalently}\qquad D_{ij,a}^{\,2}\leq cb_a\,,
 \label{eq:L-square-root-bound}
\end{equation}
where $a=\mathrm{B},\mathrm{D}.$ Both have $c=1$. For a Majorana field, $c$, $c^{\rm eff}$ and $b$ are half the Dirac values, so the entire inequality scales consistently. Independent identical copies likewise scale every term linearly. Note that the normalization coefficient in front of $\sqrt{cb}$ is not fixed by a fundamental principle, and we adopt the unit coefficient as a convenient choice.

The coefficients $r_{ij}$ in \eqref{eq:rij} are obtained by Gaussian canonical purification: diagonalizing the restricted purified correlation matrix for fermions, or its covariance symplectic spectrum for bosons, and integrating the mode entropies. At the counterexample $u=0.1$, this gives
\begin{center}
\begin{tabular}{c|cccc}
Theory & $c^{\rm eff}_{12}$ & $D_{12}$ & $b$ & $\sqrt{cb}$\\\hline
Real boson & $0.660353$ & $0.149647$ & $0.083075$ & $0.288227$\\
Dirac fermion & $0.764256$ & $0.045744$ & $0.014029$ & $0.118443$
\end{tabular}
\end{center}
The correction exceeds the deficit in both theories. At the symmetric Kirchhoff junction, the corresponding coefficients are $b_{\rm B}\simeq0.127991$ and $b_{\rm D}\simeq0.059856$, although the uncorrected bound already holds there.

The square root is important near a decoupled third leg. Along the family defined above, $u\to0^+$ gives the Dirac asymptotics
\begin{equation}
 D_{12,\rm D}\simeq \frac6{\pi^2}u\log\frac1u,\qquad
 b_{\rm D}\simeq \left(\frac4{\pi^4}\right)^{1/3}
 u^{5/3}\left(\log\frac1u\right)^{2/3}.
\end{equation}
Consequently, $D_{12,\rm D}/b_{\rm D}\to\infty$, ruling out a fixed linear correction $\kappa b$ along this path, whereas
\begin{equation}
 \frac{D_{12,\rm D}}{\sqrt{b_{\rm D}}}
 =\mathcal{O}\!\left(u^{1/6}[\log(1/u)]^{2/3}\right)\to0.
\end{equation}
For the boson, $D_{12,\rm B}\sim\tfrac32\sqrt u$ and the numerical spectral asymptotics give $b_{\rm B}\simeq0.30113\sqrt u$, hence $D_{12,\rm B}/\sqrt{b_{\rm B}}\simeq2.73u^{1/4}\to0$. Thus the square-root correction also controls this limiting counterexample in both theories. 

Exploratory scans of the real N--D junction family reported maximum ratios $D_{ij}/\sqrt{cb}$ of approximately $0.689$ for bosons and $0.8803$ for Dirac fermions, both below one. Together with the limiting checks, these results support L-entropy as a unified correction shown in \eqref{eq:L-square-root-bound} for the two free theories.

The geometric description of reflected entropy uses the entanglement wedge cross-section~\cite{TakayanagiUmemoto2018,DF}. For two-CFT interfaces, this prescription has been used to compute reflected entropy for finite regions in vacuum and thermofield-double states~\cite{BasakEtAlReflected2023}. 
For the holographic complete-leg prescription considered here, the corresponding entanglement wedge cross-sections give $S_R(i:j)=2\min\{S(\mathcal H_i),S(\mathcal H_j)\}$ for at least one pair. Thus one $\ell_{ij}$ vanishes, implying $\mathcal L_{123}=b=0$ at leading classical order. The square-root correction consequently cannot repair the holographic pairwise-bound violations in this prescription. We do not extend this conclusion to arbitrary holographic subregions. 

\section{Conclusion and discussion}
\label{sec:conclusion}

We have studied the relation between entanglement and energy transmission in conformal Y-junctions. For holographic junctions, we formulated a branched RT prescription with a common-trace homology condition and extracted pairwise effective central charges from mutual information. The resulting classical entropies satisfy subadditivity and monogamy of mutual information. Nevertheless, the pairwise effective central charge does not always bound the corresponding energy transmission. For the specified equal-length regions and two holographic radius families, we established a corrected bound, 
\begin{equation}
 c_i\mathcal T_{ij}\leq c_{ij}^{\mathrm{eff}}+3(-I_3)
 \leq\min\{c_i,c_j\}+3(-I_3),\qquad I_3\leq0.
\end{equation}
The coefficient $3$ is fixed by limiting saturation in the dominant-radius family, whereas the equal-larger-radius family has a strictly positive corrected gap. 

We also found violations of the pairwise bound in junctions of free real bosons and Dirac fermions. The holographic tripartite correction does not extend to these examples: the finite $I_3$ is positive in the bosonic realization studied here and vanishes in the continuum Dirac example. Instead, analytic limiting behavior and numerical results support a candidate correction based on the complete-leg L-entropy coefficient $b$,
\begin{equation}
 c\mathcal T_{ij}\leq c^{\rm eff}_{ij}+\sqrt{c\,b}.
\end{equation}
A general proof remains open. In the holographic complete-leg prescription considered here, $b$ vanishes, so this correction cannot restore the violated bound. 

These results show that $I_3$ and L-entropy capture different aspects of multipartite correlations, although their comparison involves different subsystem partitions and does not establish a universal distinction between holographic and free-field theories~\cite{BalasubramanianEtAl2024,LE}.


An important open question is whether a common three-party correction can bound energy transmission in Y-junctions at arbitrary coupling strength.   Schematically, the desired bound would take the form
\begin{equation}
 c_i\mathcal T_{ij}\leq c_{ij}^{\mathrm{eff}}+\Delta_{ij}^{(3)},
 \qquad \Delta_{ij}^{(3)}\geq0,
 \label{eq:bound-delta}
\end{equation}
where $\Delta_{ij}^{(3)}$ remains to be constructed. Such a quantity should be defined with a consistent partition and regulator, have the same normalization in both descriptions, and remain effective when $I_3$ changes sign or vanishes and when the L-entropy coefficient vanishes. It should also recover the two-party interface bound when the third leg decouples and scale consistently under independent copies. 
Note that the right side of \eqref{eq:bound-delta} is further constrained by $\min\{c_i,c_j\}+\Delta_{ij}^{(3)}.$

A natural candidate that accommodates both corrections is
\begin{equation}
 \Delta_{ij}^{(3)}=3|I_3|+\sqrt{c_i\,b}.
\end{equation}
This expression reproduces the holographic correction and is consistent with 
the free-field junctions. It would be interesting to determine whether this combined correction, with an appropriate generalization to unequal central charges, satisfies~\eqref{eq:bound-delta} for general conformal Y-junctions.  
Another possible candidate for $\Delta_{ij}^{(3)}$ might be constructed from replica-based multi-entropy, which is symmetric under permutations of the parties and has a geometric prescription in holography~\cite{GaddeKrishnaSharma2022, Penington:2022dhr, Ju:2025eyn, Iizuka:2025ioc}. A concrete test is whether a contribution derived from this measure can bound the channel deficit and vanish when the third leg decouples. Further tests include proving or delimiting the L-entropy bound for weakly coupled Y-junction, extending the holographic analysis to unequal intervals and additional RT phases, and studying junctions with more than three legs.


It would also be interesting to extend the analysis beyond conformal junctions. Brane-localized dynamics can generate interface RG flows and nontrivial finite-temperature geometries~\cite{LiuLyuWang2024}, while non-conformal interfaces exhibit frequency-dependent energy transport~\cite{LiuWangZeng2025, Ge:2026elr}. Comparing these transport coefficients with entanglement requires identifying appropriate information measures and scales. 
Stationary thin-brane solutions describing heat currents across interfaces~\cite{BachasChenPapadopoulos2021} provide a starting point for extending the analysis to steady heat flow through junctions. Finite-temperature and time-dependent states would further test the applicability of these bounds beyond pure ground states.


\subsection*{Acknowledgments}
We thank Ling-Long Gao, Libo Jiang, Ya-Wen Sun, Chuan-Yi Wang, Ren-Jie Ying and Shao-Dong Zhou for useful discussions. This work was supported by the National Natural Science Foundation of China Grants No. 12375041 and 12575046.

\appendix
\section{The information measures from the branched RT prescription}
\label{app:junction-information}

In this appendix, we collect the geodesic lengths needed for the branched RT prescription, derive the leading mutual-information coefficients, and evaluate the finite tripartite information in the two holographic radius families. We use the regulated regions defined in~\eqref{eq:junction-i3-regions} and minimize total RT length over configurations satisfying the
common-trace homology condition of Sec.~\ref{subsec:rt-networks}.

\subsection{Geodesic lengths and mutual information}
\label{app:mutual-information}


On a constant-time slice, the geodesic distance between two points $P=(x,u)$ and $P'=(x',u')$ in branch $r$ is~\cite{AnousEtAl2022}
\begin{equation}
 \mathcal D_r(P,P')=L_r\operatorname{arcosh}
 \frac{(x-x')^2+u^2+u'^2}{2uu'}.
 \label{eq:mi-general-geodesic}
\end{equation}
We regulate a boundary endpoint at $E_r(x)=(x,\epsilon\cos\theta_r)$ and write $P_r(z)$ for the embedding in branch $r$ of the point on $Q$ with intrinsic coordinate $z$, as in~\eqref{eq:hj-embedding}. We then obtain the boundary-to-brane, boundary-to-boundary, and brane-to-brane kernels,
\begin{align}
 \begin{split}
 D_r(x,z)
 &=L_r\operatorname{arcosh}
 \frac{x^2-2xz\sin\theta_r+z^2+\epsilon^2\cos^2\theta_r}
 {2\epsilon z\cos^2\theta_r},
 \\
 H_r(x,x')
 &=L_r\operatorname{arcosh}\left(1+\frac{(x-x')^2}{2\epsilon^2\cos^2\theta_r}\right),
 \\
 K_r(z_-,z_+)
 &=L_r\operatorname{arcosh}
 \frac{z_-^2+z_+^2-2z_-z_+\sin^2\theta_r}
 {2z_-z_+\cos^2\theta_r}.
 \label{eq:mi-geodesic-kernels}
  \end{split}
\end{align}
These are distances in the ambient AdS$_3$ geometry. We minimize the total RT length over configurations whose geodesic
segments lie within their respective bulk branches and satisfy the junction homology condition.

For the one-vertex candidates associated with $A_i$ and $A_iA_j$, we minimize
\begin{align}
\begin{split}
 F_i&=\min_{z>0}[D_i(-\ell,z)+D_j(-\delta,z)+D_k(-\delta,z)],\\
 F_{ij}&=\min_{z>0}[D_i(-\ell,z)+D_j(-\ell,z)+D_k(-\delta,z)].
 \label{eq:mi-Y-lengths}
 \end{split}
\end{align}
For $A_j$, the disconnected and two-vertex candidates are
\begin{align}
\begin{split}
 F_j^{\mathrm{disc}}&=H_j(-\delta,-\ell),\\
 F_j^{\mathrm{br}}&=\min_{0<z_-<z_+}
 [D_j(-\delta,z_-)+D_j(-\ell,z_+)+K_i(z_-,z_+)+K_k(z_-,z_+)].
 \label{eq:mi-j-lengths}
  \end{split}
\end{align}
In a phase where these candidates dominate, the mutual information is explicitly
\begin{equation}
 I(A_i:A_j)=\frac{F_i+\min(F_j^{\mathrm{disc}},F_j^{\mathrm{br}})-F_{ij}}{4G_N}.
 \label{eq:mi-finite-expression}
\end{equation}

The same subregions and node assignment are used in all three entropy terms. Each of the three entropies must be minimized separately: its shortest RT surface may be connected or disconnected, depending on the radii and brane angles.

To extract the logarithmic terms, let $W=\log(\ell/\epsilon)$. A segment from a macroscopic boundary endpoint to a vertex with $z=\mathcal{O}(\ell)$ contributes $L_r W+\mathcal{O}(1)$. A segment connecting a vertex with $z=\mathcal{O}(\epsilon)$ to one with $z=\mathcal{O}(\ell)$ contributes the same coefficient. A direct segment between $-\delta$ and $-\ell$ contributes $2L_rW+\mathcal{O}(1)$, and segments whose coordinates all scale with $\epsilon$ contribute only $\mathcal{O}(1)$. These statements follow by expanding the kernels in 
\eqref{eq:mi-geodesic-kernels} 
at fixed $\lambda$.

For $A_i$ or $A_j$, the outer endpoint supplies a logarithm on its own branch. The node cut can run along that branch or along the two complementary branches. For $A_iA_j$, there are two outer endpoints and the node cut separates the complementary branch $k$ from $i,j$. Minimizing the logarithmic cost gives $h_r$ in~\eqref{eq:mi-single-cut-charge}. Using $L_r/(4G_N)=c_r/6$, we obtain
\begin{align}
\begin{split}
 S(A_i)&=\frac{c_i+h_i}{6}\log\frac{\ell}{\epsilon}+\mathcal{O}(1),\\
 S(A_j)&=\frac{c_j+h_j}{6}\log\frac{\ell}{\epsilon}+\mathcal{O}(1),\\
 S(A_iA_j)&=\frac{c_i+c_j+h_k}{6}\log\frac{\ell}{\epsilon}+\mathcal{O}(1).
 \label{eq:mi-leading-entropies}
  \end{split}
\end{align}
Combining these three entropies gives~\eqref{eq:mi-leading}.  The remaining finite terms may depend on the brane angles and the fixed regulator ratio $\lambda$.

\subsection{Finite tripartite information}
\label{app:tripartite-information}

We now extract the finite part of $I_3$ for the same regulated partition. 
In branch $i$, $D_i(x,z)$ is the geodesic length from a regulated boundary endpoint at $x$ to a brane point at $z$, $H_i(x,x')$ joins two regulated boundary endpoints, and $K_i(z_-,z_+)$ joins two brane points. Their explicit expressions are given in  
\eqref{eq:mi-geodesic-kernels}. Write $s_i=\sin\theta_i$. For a macroscopic brane vertex $z$, their UV expansions give
\begin{align}
\begin{split}
 D_i(-\ell,z)&=L_i\log\frac{L_a^2(z^2+2s_i\ell z+\ell^2)}{L_i^2z\epsilon}+\mathcal{O}(\epsilon),\\
 B_i&:=H_i(-\delta,-\ell)=L_i\log\frac{L_a^2\ell^2}{L_i^2\epsilon^2}+\mathcal{O}(\epsilon),\\
 K_i(\epsilon u,z)&=g_i(z)-L_i\log u+\mathcal{O}(\epsilon),\qquad
 g_i(z)=L_i\log\frac{L_a^2z}{L_i^2\epsilon}.
 \label{eq:junction-i3-geodesic-expansions}
\end{split}
\end{align}
Here $\epsilon u$ is a near-node vertex coordinate, while $z=\mathcal{O}(\ell)$. 

\begin{figure}[h!]
\centering
\begin{tikzpicture}[x=.88cm,y=.88cm,font=\small,line cap=round,line join=round,
 far/.style={blue!75!black,line width=1.4pt},
 near/.style={orange!85!black,dashed,line width=1.2pt},
 bridge/.style={violet!80!black,line width=1.4pt},
 direct/.style={teal!75!black,line width=1.4pt}]
\begin{scope}
\coordinate (O) at (0,0);\coordinate (P) at (0,3.1);\coordinate (R) at (0,.85);
\coordinate (U) at (3,1.2);\coordinate (V) at (3,-1);
\draw[gray!75] (-2.7,0)--(O)--(3.5,1.4);\draw[gray!75] (O)--(3.5,-1.17);
\draw[gray!85] (O)--(0,3.6) node[above] {$Q$};
\draw[far] (U)..controls (2.65,2.7) and (1.3,3.1)..(P);
\draw[far] (V)..controls (2.9,.8) and (1.3,2.9)..(P);
\draw[near] (.6,.24)..controls (.65,.65) and (.35,.85)..(R);
\draw[near] (.6,-.2)..controls (.6,.3) and (.3,.75)..(R);
\draw[bridge] (R)..controls (-2,.95) and (-2,2.9)..(P);
\fill (P) circle (1.8pt);\fill (R) circle (1.8pt);
\node[above right] at (P) {$z$};\node[above left] at (R) {$\epsilon u$};
\node[blue!75!black,above] at (1.85,2.8) {$D_2(-\ell,z)$};
\node[blue!75!black,right] at (2.3,.7) {$D_3(-\ell,z)$};
\node[violet!80!black,left,align=center] at (-1.7,2) {$K_1(\epsilon u,z)$};
\node[below] at (-2.1,0) {$\mathrm{CFT}_1$};

\node[below] at (.6,-.2) {$-\delta$};
\node[above] at (3.5,1.4) {$\mathrm{CFT}_2$};
\node[below] at (3.5,-1.17) {$\mathrm{CFT}_3$};
\node[align=center] at (.5,-1.95) {(a) $T=\{2,3\}$: $D_2+D_3+g_1$};
\end{scope}
\begin{scope}[xshift=8.2cm]
\draw[gray!75] (-1.7,0)--(0,0)--(3.5,1.4);\draw[gray!75] (0,0)--(3.5,-1.17);
\draw[gray!85] (0,0)--(0,3.6) node[above] {$Q$};
\draw[direct] (.6,.24)..controls (.45,2.65) and (2.15,3.05)..(3,1.2);
\node[teal!75!black,above,align=center] at (1.6,2.5) {$H_2(-\delta,-\ell)$};
\node[below] at (.6,.24) {$-\delta$};\node[below right] at (3,1.2) {$-\ell$};
\node[above] at (3.5,1.4) {$\mathrm{CFT}_2$};
\node at (1,-1.95) {(b) $B_2=H_2(-\delta,-\ell)$};
\end{scope}
\end{tikzpicture}
\caption{\small Geodesic segments for $L_1\leq L_2<L_3$ with $L_1+L_2<L_3$. Blue: outer-endpoint lengths $D_i$. Violet: $K_1$, whose far part is $g_1$. Dashed orange: boundary-to-brane segments entering the near-node contributions together with the near part of $K_1$. Teal: the direct interval length $B_2$. Curves are schematic.}
\label{fig:i3-length-segments}
\end{figure}
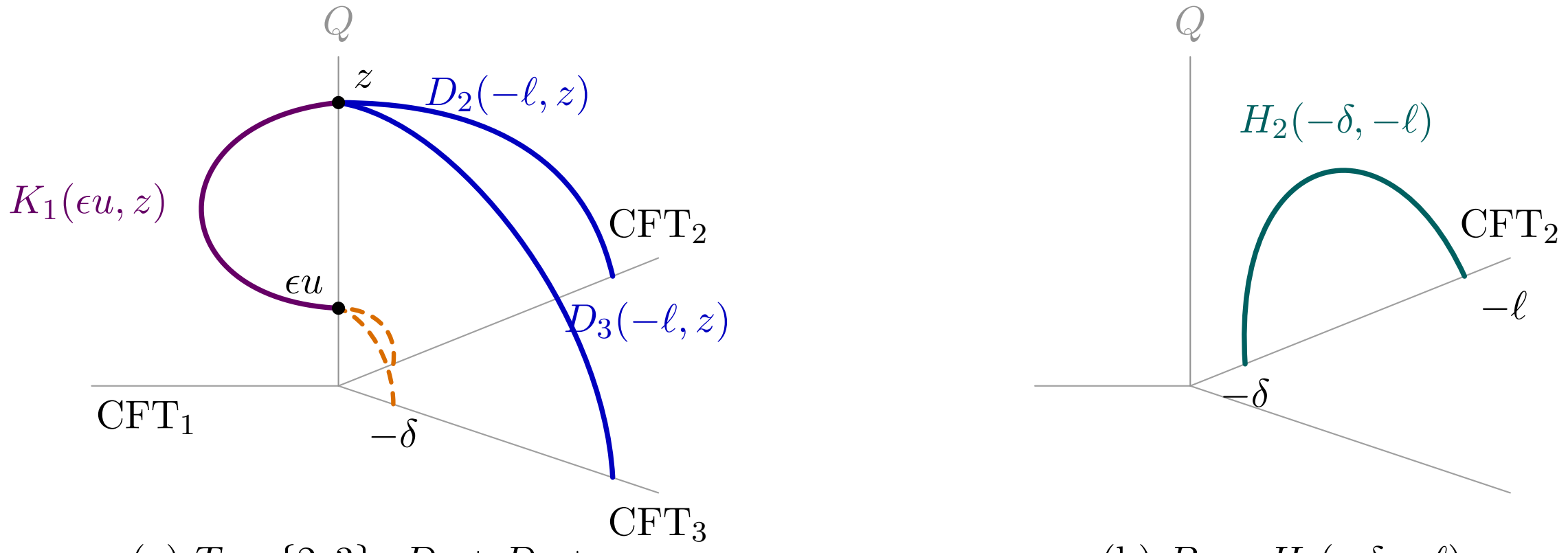

We obtain the macroscopic contribution $\mathcal F_T$ by minimizing the sum of the segment lengths over the common vertex position $z$.
\begin{equation}
 \mathcal F_T=\min_{z>0}\left[
 \sum_{i\in T}D_i(-\ell,z)+\sum_{i\notin T}g_i(z)\right].
 \label{eq:junction-i3-macroscopic}
\end{equation}
Set $T\subseteq\{1,2,3\}$ labels the branches with macroscopic outer endpoints, and $i\notin T$ denotes its complement. Each branch in $T$ connects its outer endpoint to the vertex, and the other branches contribute the macroscopic parts of their brane-to-brane segments.
The minimized near-node contributions, formed from the dashed orange segments in Fig.~\ref{fig:i3-length-segments} and the near part of $K_1$ in~\eqref{eq:junction-i3-geodesic-expansions}, are shared by complementary entropy terms and cancel in $I_3$. 

For each of the seven entropy terms, we minimize the total RT length over all admissible configurations, including both connected and disconnected prescription. Combining the minimizing lengths in each radius family gives
\begin{equation}
 J=\begin{cases}
 \mathcal F_{ijk}-\mathcal F_{ik}-\mathcal F_{jk}+\mathcal F_k,
 &L_i\leq L_j<L_k,\ L_i+L_j<L_k,\\[3pt]
 B_i+B_j+B_k+\mathcal F_{ijk}-\mathcal F_{ij}-\mathcal F_{ik}-\mathcal F_{jk},
 &L_i\leq L_j=L_k.
 \end{cases}
 \label{eq:junction-i3-macro-combinations}
\end{equation}
Here $\{i,j,k\}=\{1,2,3\}$ follows the subregion notation in~\eqref{eq:junction-i3-regions}. In the first row, $k$ labels the dominant-radius branch. The logarithmic divergences in~\eqref{eq:junction-i3-geodesic-expansions} cancel in both combinations of~\eqref{eq:junction-i3-macro-combinations}, leaving only the finite parts as the $\mathcal{O}(\epsilon)$ corrections vanish for $\epsilon \to 0$. 

The brane angles and $L_a$ are fixed by the tension relation~\eqref{eq:hj-tension}. The two geometric branches in~\eqref{eq:transport-dominant-branches} and the positive-angle family specified in~\eqref{eq:transport-equal-minimum} therefore determine the dependence of $I_3$ on $\tau$ and the AdS radii. To evaluate the macroscopic terms, set $z=\ell e^{-t}$ and let $C_T$ be the value of the length functional in~\eqref{eq:junction-i3-macroscopic} at $z=\ell$. The boundary-to-brane lengths differ from their values at $z=\ell$ by $L_r\log[(\cosh t+s_r)/(1+s_r)]$, while $g_r(z)-g_r(\ell)=-L_rt$. The logarithm involving $\cosh t$ is even, so a minimum can be chosen at $t\geq0$. Thus the UV expression is
\begin{equation}
 \mathcal F_T=C_T-E_T+\mathcal{O}(\epsilon),\qquad
 E_T=\sup_{t\geq0}\left[
 t\sum_{r\notin T}L_r-\sum_{r\in T}L_r\log\frac{\cosh t+s_r}{1+s_r}
 \right].
\end{equation}
In particular, $E_{123}=0$. The remaining maximizations give the two cases below.

\paragraph{The dominant-radius family: $L_1\leq L_2<L_3$ with $L_1+L_2<L_3$.}
Put $M=L_1+L_2$ and define
\begin{equation}
 \phi_i(t)=\log\frac{\cosh t+s_i}{1+s_i},\qquad t\geq0.
 \label{eq:transport-phi}
\end{equation}
The baseline lengths cancel because $C_{123}-C_{13}-C_{23}+C_3=0$, and we obtain
\begin{align}
\begin{split}
 -J&=E_3-E_{13}-E_{23},\\
 E_{13}&=\sup_{t\geq0}\{L_2t-L_1\phi_1(t)-L_3\phi_3(t)\},\\
 E_{23}&=\sup_{t\geq0}\{L_1t-L_2\phi_2(t)-L_3\phi_3(t)\},\\
 E_3&=\sup_{t\geq0}\{Mt-L_3\phi_3(t)\}.
 \label{eq:transport-dominant-variational}
 \end{split}
\end{align}

The last maximization has an explicit elementary solution. With $p=M$, $L=L_3$ and $s=s_3$, set
\begin{equation}
 v_p=\frac{ps+\sqrt{L^2-p^2(1-s^2)}}{L-p},\qquad
 E_3=p\log v_p-L\log\frac{(v_p+v_p^{-1})/2+s}{1+s}.
 \label{eq:junction-i3-explicit-vertex}
\end{equation}
The other two maxima are fixed by $L_1\phi_1'(t)+L_3\phi_3'(t)=L_2$ and $L_2\phi_2'(t)+L_3\phi_3'(t)=L_1$, respectively, with the maximizing solutions selected. These equations become algebraic after setting $v=e^t$. Equal interval lengths therefore simplify the calculation, but do not by themselves give a single universal elementary formula for arbitrary radii and angles.

\paragraph{The equal-larger-radius family: $L_1=a\leq L_2=L_3=b$.}
Set $a=L_1$, $b=L_2=L_3$ and $s=s_2=s_3>0$ on the positive-angle branch, with $\phi_a=\phi_1$ and $\phi_b=\phi_2=\phi_3$. The two maxima are
\begin{equation}
 E_1=\sup_{t\geq0}\{at-2b\phi_b(t)\},\qquad
 E_2=\sup_{t\geq0}\{bt-a\phi_a(t)-b\phi_b(t)\}.
 \label{eq:transport-equal-variational}
\end{equation}
We obtain $E_1$ from~\eqref{eq:junction-i3-explicit-vertex} 
by taking $(p,L,s)=(a,2b,s)$; the second is determined by $a\phi_a'(t)+b\phi_b'(t)=b$. In the second row of~\eqref{eq:junction-i3-macro-combinations}, the baseline terms sum to $\sum_r[B_r-D_r(-\ell,\ell)-g_r(\ell)]$. For each branch, the expansions~\eqref{eq:junction-i3-geodesic-expansions} and $L_r^2/L_a^2=1-s_r^2$ reduce the bracket to $L_r\log[(1-s_r)/2]+\mathcal{O}(\epsilon)$. The finite tripartite combination becomes
\begin{equation}
 J=E_1+2E_2
   +a\log\frac{1-s_1}{2}+2b\log\frac{1-s}{2}.
 \label{eq:transport-equal-i3}
\end{equation}

The stationary equations specifying the $E$'s become algebraic after $x=e^{-t}$ is introduced. Selecting their global maximizing solutions gives an implicit analytic expression without requiring lengthy radical formulas. Together with the tension relation~\eqref{eq:hj-tension}, the resulting entropy is
\begin{equation}
 I_3(\tau;L_1,L_2,L_3)=\frac{J}{4G_N}.
 \label{eq:transport-i3-normalization}
\end{equation}
Within the RT curves considered here, the finite result is independent of the interval length $\ell$, the UV cutoff $\epsilon$, and the fixed ratio $\lambda$. 


\end{document}